\documentclass[preprint,12pt]{elsarticle}

\def\tabfont{\fontsize{10}{10}\selectfont}

\biboptions{sort&compress}
\usepackage{graphicx, subfigure}
\usepackage{longtable}
\usepackage{lscape}
\usepackage{epsfig}
\usepackage{amssymb}
\usepackage{amsmath}

\usepackage{placeins}
\usepackage{multirow}
\usepackage{multirow}
\usepackage[flushleft]{threeparttable}
\usepackage{array}
\newcommand{\head}[2]{\multicolumn{1}{>{\centering\arraybackslash}p{#1}}{#2}}
\usepackage{rotating}
\usepackage{chngpage}
\usepackage{booktabs}
\usepackage{geometry}

\usepackage{etoolbox}
\patchcmd{\pprintMaketitle}
 {\ifvoid\absbox\else\unvbox\absbox\par\vskip10pt\fi}
 {\ifvoid\absbox\else\clearpage\unvbox\absbox\par\vskip30pt\fi}
 {}{}
\patchcmd{\pprintMaketitle}
 {\hrule\vskip12pt}
 {}
 {}{}
\patchcmd{\pprintMaketitle}
 {\hrule\vskip12pt}
 {}
 {}{}
\appto{\pprintMaketitle}{\clearpage}

\usepackage{lineno}
\usepackage{lineno}
\usepackage{multirow}
\usepackage{booktabs}
\usepackage{amsbsy}
\usepackage{soul}
\usepackage{color}

\biboptions{comma,compress}

\journal{ApJ}

\begin{document}

\begin{frontmatter}

\title{Searches for Time Dependent Neutrino Sources with IceCube Data from 2008 to 2012}

\author[Adelaide]{M.~G.~Aartsen}
\author[Zeuthen]{M.~Ackermann}
\author[Christchurch]{J.~Adams}
\author[BrusselsLibre]{J.~A.~Aguilar}
\author[MadisonPAC]{M.~Ahlers}
\author[StockholmOKC]{M.~Ahrens}
\author[Erlangen]{D.~Altmann}
\author[PennPhys]{T.~Anderson}
\author[MadisonPAC]{C.~Arguelles}
\author[PennPhys]{T.~C.~Arlen}
% ...
\address[Adelaide]{School of Chemistry \& Physics, University of Adelaide, Adelaide SA, 5005 Australia}
% ...
\address[BrusselsLibre]{Universit\'e Libre de Bruxelles, Science Faculty CP230, B-1050 Brussels, Belgium}
% ...
\address[Christchurch]{Dept.~of Physics and Astronomy, University of Canterbury, Private Bag 4800, Christchurch, New Zealand}
% ...
\address[Erlangen]{Erlangen Centre for Astroparticle Physics, Friedrich-Alexander-Universit\"at Erlangen-N\"urnberg, D-91058 Erlangen, Germany}
% ...
\address[MadisonPAC]{Dept.~of Physics and Wisconsin IceCube Particle Astrophysics Center, University of Wisconsin, Madison, WI 53706, USA}
% ...
\address[StockholmOKC]{Oskar Klein Centre and Dept.~of Physics, Stockholm University, SE-10691 Stockholm, Sweden}
% ...
\address[PennPhys]{Dept.~of Physics, Pennsylvania State University, University Park, PA 16802, USA}
% ...
\address[Zeuthen]{DESY, D-15735 Zeuthen, Germany}
\fntext[Tokyofn]{Earthquake Research Institute, University of Tokyo, Bunkyo, Tokyo 113-0032, Japan}
\fntext[Goddard]{NASA Goddard Space Flight Center, Greenbelt, MD 20771, USA}

\begin{abstract}
In this paper searches for flaring astrophysical neutrino sources and sources with periodic emission with the IceCube neutrino telescope are presented. In contrast to time integrated searches, where steady emission is assumed, the analyses presented here look for a time dependent signal of neutrinos using the information from the neutrino arrival times to enhance the discovery potential. A search was performed for correlations between neutrino arrival times and directions as well as neutrino emission following time dependent lightcurves, sporadic emission or periodicities of candidate sources.
These include active galactic nuclei, soft $\gamma$-ray repeaters, supernova remnants hosting pulsars, micro-quasars and X-ray binaries. The work presented here updates and extends previously published results to a longer period that covers four years of data from 2008 April 5 to 2012 May 16 including the first year of operation of the completed 86-string detector. The analyses did not find any significant time dependent point sources of neutrinos and the results were used to set upper limits on the neutrino flux from source candidates.
\end{abstract}

\begin{keyword}
triggered searches, multi-messenger searches, multi-wavelength campaigns, blazars, active galaxies, $\gamma$-ray bursts, 
soft $\gamma$-ray repeaters, X-ray binaries\end{keyword}                   

\end{frontmatter}
\clearpage
\section{Introduction} 
\label{intro}

The cosmic ray spectrum spans ten decades in energy up to $10^{11}$~GeV per particle and despite being extensively studied for many years the origin and the acceleration mechanism remains uncertain. Cosmic rays consist of hadrons, mainly protons and in part also ionized nuclei with an energy dependent composition. As a result of interactions of cosmic rays with matter and ambient photons close to the acceleration sites pions are produced and decays of the charged pions and their daughter muons then produce neutrinos. Such astrophysical neutrinos are a unique and valuable messenger in astro-particle physics because of their properties and the specifics of their production mechanisms. In particular, they carry information about the origin and spectrum of cosmic rays. In contrast to the cosmic rays, neutrinos are not deflected in magnetic fields, nor do they interact on the way to Earth. As a consequence their trajectories point back to their origin. A detection of neutrinos from a given site would 
therefore stand as proof of accelerated hadrons, identifying it as a source of cosmic rays.

This paper updates and expands the IceCube time dependent searches and results for flaring sources~\cite{mike, ic59-time, crab} and periodic sources~\cite{periodical}. Time dependent astrophysical neutrino signals can be better observed by using event times in addition to the direction and energy used in standard IceCube point source searches, because the additional information improves the rejection of the atmospheric muon and neutrino events which form the dominant background.

If several events are coming from the same astrophysical point source they should be spatially concentrated around the emitting source and they should have a harder spectrum than muons and neutrinos produced in atmospheric showers. Assuming a Fermi acceleration model~\cite{FermiAcc,FermiAcc2} a differential neutrino spectrum close to~$E^{-2}$ will be produced. Additional effects like the acceleration of muons in the cosmic ray sources~\cite{mu_acc1,mu_acc2} can modify the spectrum. In case of sufficiently high acceleration gradients (above 160 keV/m) often required for extremly short flares, the daughter muons from charged pion decays can be accelerated before they decay. Thus the energy of the neutrinos from the muon decays will be enhanced. The impact on the overall neutrino spectra strongly depends on various properties of the source, e.g. the magnetic field strength and the amount of matter in the acceleration region. Conventional atmospheric neutrinos at energies above 100 GeV have a much softer 
spectrum which asymptotically approaches a spectrum one power steeper than the primary spectrum, first in the vertical direction and at higher energies also for larger zenith angles~\cite{ic59_diffuse,nufluxes}. 

Time dependent searches can be performed in an `untriggered' way, meaning that the correlation of event times is investigated and no additional information from independent observations is used. The full parameter space of time, energy and direction of measured events is scanned looking for clusters in time and space of high energy events among the background of atmospheric events, hereby called the ``All-Sky Time Scan''.
This search is the most generic one but it is subject to a large trial factor that penalizes the significance of the signal with respect to the background. Hence, in addition, more specific searches are carried out using a multi-messenger approach. The basic assumption is that neutrinos and $\gamma$-rays are correlated because they have a common origin in the astrophysical sources that accelerate the cosmic rays. 
The ``Search for Triggered Multi-Messenger Flares'' makes use of lightcurves measured by $\gamma$-ray experiments. This analysis is triggered by multi-wavelength measurements (alerts for flares from TeV and X-ray experiments)
as well as the Fermi-LAT lightcurves which provide continuous monitoring of selected sources or of flares above a certain photon flux level~\cite{Fmonitored}.

The paper outline is as follows. In Section~\ref{sec:sources} the possible sources of astrophysical neutrinos are briefly discussed and in Section~\ref{sec:detsamp} the IceCube detector and the details of the data samples used are introduced. Section~\ref{sec:llh} gives a detailed description of the general likelihood method applied while the specifics, which are different for each search, are given at the beginning of the corresponding sections.

In the following sections, the results for the various searches are presented. First in Section~\ref{sec:All-Sky Time Scan} the untriggered ``All-Sky Time Scan'' looking for any cluster of high energy neutrino events from any direction in the sky is presented. For this search the data taken by the 59-string configuration of IceCube (IC-59), 79-string configuration (IC-79) and the first year of the full IceCube data taking with 86 strings (IC-86I) were used. For each of the three years time dependent skymap scans were performed.
 
Then in Section~\ref{sec:fflares} the results of the ``Search for Triggered Multi-Messenger Flares'' are discussed. The data taken with the IC-59, IC-79 and IC-86I configurations of IceCube were combined. Hence, this represents a long term study for flaring objects. In Section~\ref{sec:sflares} the results of the ``Search for Triggered Flares with Sporadic Coverage'' are presented. These were triggered by higher energy~$\gamma$-ray observations in the~TeV~range.

In Section~\ref{sec:per} the ``Search for Periodic Neutrino Emissions from Binary Systems'' is presented. The search was performed in the phase domain. For this search data from four IceCube data taking seasons, from the IC-40 to the IC-86I  configuration were used.

Finally, in Section~\ref{sec:Conclusions} we conclude with a brief outlook.

\section{Potential Flaring Sources of Neutrinos}
\label{sec:sources}

The focus of this paper is on active galactic nuclei (AGNs), and particularly blazars, i.e. AGNs with jets pointing toward us, which are interesting due to their variability and the high power emitted during bursts~\cite{AGNs1,AGNs2,AGNs3}. The analyses described in the following sections are sensitive also to short flares (of the duration of a few seconds) and hence, to some extent, to $\gamma$-ray bursts. More sensitive searches for neutrinos from $\gamma$-ray bursts are presented elsewhere \cite{naturepaper,grb_ic40}. Blazars exhibit sudden sequences of multiple flares that may last from minutes to months and are observed in various wavelengths from radio to $\gamma$-rays. Correlations between various bands have been observed in numerous multi-wavelength campaigns, particularly between X-rays and~$\gamma$ TeV emissions. A well-studied case is that of the close-by TeV blazar Mrk 421 (see e.g. \cite{mrk2010,magic_mrk421}). In other instances optical flares have triggered TeV flare observations \cite{optical}
.

The Spectral Energy Distribution (SED) of AGNs is characterized by two broad peaks. The lower energy one is believed to originate from synchrotron emission of the charged particles in
the jet. In leptonic models the higher energy one is generally explained by inverse Compton scattering of either the synchrotron seed photons (Synchrotron Self Compton - SSC, see e.g. \cite{blazar1}) or external seed photons (External Compton - EC \cite{blazar2,blazar3}) by the electrons and positrons in the jet. 

In the simplest case, both SSC and EC mechanisms predict that flaring at TeV energy should be accompanied by a simultaneous flaring in the synchrotron peak and so a connection between bands from optical to $\gamma$-ray is expected. If the synchrotron peak is located far from the optical, as in the case of High-frequency Peaked Blazars (HBLs), then the synchrotron flares should be visible at other wavelengths, usually X-rays \cite{mrk2010,magic_mrk421}. 

Alternatively, hadronic models suggest production of neutrinos and $\gamma$-rays from pion decays~\cite{blazar4,blazar5,blazar6}. In these models, the high energy peak is due to proton synchrotron emission or decay of neutral pions formed in cascades by the interaction of high-energy proton beam with the radiation or gas clouds surrounding the source \cite{blazar6}. In this scenario, a strong correlation between the $\gamma$-ray and the neutrino fluxes is expected.
Observations of neutrinos would clearly distinguish between leptonic and hadronic models.

For some observations it has been claimed that hadronic processes could explain flares better than leptonic processes~\cite{reimer}. Orphan flares, i.e. a TeV flare without a lower energy counterpart, challenge leptonic models~\cite{orphan1,orphan2,orphan3}. Non-observation of significant X-ray activity could naturally be interpreted as due to the suppression of electron acceleration and inverse Compton scattering and dominance of very high energy (VHE) $\gamma$-ray production from meson decays in hadronic models.

In addition to flares, another possible time dependent signature is a periodicity in the neutrino emission. In the case of binary systems, a modulation of the neutrino emission could occur due to the relative geometry of the emitted beam with respect to the large mass star and the observer. A particularly interesting case is that of micro-quasars, which are radio-jet X-ray binaries that can include either a neutron star or a black hole. A similar modulation is observed for three binary systems in TeV $\gamma$-rays: LS 5039 \cite{mqso1}, LS I +61 303 \cite{mqso2}, and HESS J0632+057 \cite{mqso3}. These three binary systems are found to be periodic at GeV and TeV energies, although the emission in the two bands seems anti-correlated \cite{mqso_anti}. This anti-correlation is in fact a generic feature of models in which the GeV (TeV) emission is enhanced (reduced) when the highly relativistic electrons are moving in the direction of the observer and encounter the seed photons head-on. Neutrino emission from
micro-quasars has been described in various papers \cite{mqso_nu1,mqso_nu2,mqso_nu3,mqso_nu4,mqso_nu5}. It can be assumed a neutrino signal would be correlated to the periodic $\gamma$-ray emission from the binary system due to the system's rotation, but it is not clear in which phase with respect to the $\gamma$-rays the neutrinos are to be expected. Hence the values for the period are taken from multi-wavelength observations and the phase is fitted as a free parameter.

\section{The IceCube Detector and the Data Samples}
\label{sec:detsamp}

The IceCube observatory is a Cherenkov detector searching for high energy neutrinos. It is an array of 5,160 Digital Optical Modules (DOMs) deployed deep in the Antarctic ice near the South Pole. The purpose of the detector is to allow observations of neutrinos of astrophysical origin and atmospheric muons and neutrinos induced by cosmic rays at energies around and above the \textit{knee} ($\mathrm{\sim 3 \times 10^{15}\:eV}$). Cherenkov light produced by secondary leptons from neutrino interactions in the vicinity of the detector is used to indirectly detect these neutrinos. Only charged-current muon neutrino events were considered for the studies presented here, because of the large range of the secondary muons, which provides good angular resolution and a high rate by including events that start outside the detector. The pointing information relies on the secondary muon direction, which at energies above TeV differs from the original 
neutrino direction by less than the angular resolution of the detector~\cite{jakerameez}.

The DOMs are spherical, pressure resistant glass housings, each containing a Hamamatsu photomultiplier tube (PMT) of 25 cm diameter and the associated electronics necessary for waveform digitization~\cite{PMT,PMTel}. These DOMs are connected together in 86 vertical strings of 60 DOMs each and the strings are lowered into 60~cm wide holes drilled into the ice using hot water to instrument a kilometer-cubed volume in a depth range from 1.5 km down to 2.5 km. At the center of the detector the strings are placed closer to each other, forming a sub-array called DeepCore. This sub-array enhances the sensitivity of the detector to neutrinos of energies below the standard IceCube threshold of~100~GeV~\cite{DeepCore}. 


The IceCube detector has been in operation even before full completion in December 2010. The partial detector layouts (resp. data samples) used for this paper are labeled IC-40, IC-59, IC-79 and IC-86I, where the number corresponds to the number of operational stings. The data samples used in the studies presented here are described in detail in previous papers reporting time integrated searches~\cite{ic40PS,juanan,jakerameez}.

\begin{table*}[t!]
\begin{center}
\vspace{0.2cm}
\small{
\begin{tabular}{lllllll}
\toprule
\head{1.0cm}{sample} &\head{1.0cm}{start (MJD)} &\head{1.0cm}{end (MJD)} &\head{1.0cm}{livetime [days]} &\head{1.0cm}{atm. $\nu$s/day} &\head{1.0cm}{up-going} &\head{1.0cm}{down-going}\\
\midrule
IC-40&\vtop{\hbox{\strut 2008 Apr 5}\hbox{\strut (54561)}} &\vtop{\hbox{\strut 2009 May 20}\hbox{\strut (54971)}} &376 &40 & 14121 & 22779 \\
IC-59&\vtop{\hbox{\strut 2009 May 20}\hbox{\strut (54971)}} &\vtop{\hbox{\strut 2010 May 31}\hbox{\strut (55347)}} &348 &120& 43339 & 64230 \\
IC-79&\vtop{\hbox{\strut 2010 May 31}\hbox{\strut (55347)}} &\vtop{\hbox{\strut 2011 May 13}\hbox{\strut (55694)}} &316 &180& 50857 & 59009 \\
IC-86I&\vtop{\hbox{\strut 2011 May 13}\hbox{\strut (55694)}} &\vtop{\hbox{\strut 2012 May 16}\hbox{\strut (56063)}} &333 & 210 & 69227 & 69095\\
\bottomrule
\end{tabular}}
  \caption{Summary of the data used in the reported time dependent point source searches. As livetime the duration of the actual data-taking is used. The fifth column, atmospheric neutrinos per day, is the expectation from Monte Carlo for the whole sky. The last two columns give the number of selected events separately for up-going and down-going for each detector configuration.}
    \label{tab:livetimes}
\end{center}
\end{table*}

The analyses in the present paper used identical detector simulation and corresponding estimates of neutrino effective area and point spread function based on Monte Carlo studies as for the time integrated analyses~\cite{jakerameez}. Table~\ref{tab:livetimes} summarizes the four data samples consisting of both up- and down-going muons which have different origin; while the up-going muons are mostly from interactions of atmospheric neutrinos that have passed through the Earth, the down-going muons are mainly from meson decays in atmospheric showers caused by cosmic rays.

\section{Unbinned Time Dependent Likelihood Method}
\label{sec:llh}
The unbinned time dependent likelihood method which was used for the analyses in this paper has been applied in previous analyses~\cite{method1} and~\cite{method2}. 

The likelihood function is defined as:

\begin{equation}
\label{eq:likelihood}
\mathcal{L}(\gamma, n_{\mathrm{s}},...) = \prod_{j} \mathcal{L}^{j} (\gamma, n^{j}_{\mathrm{s}},...) = \prod_{j}\;\prod_{i \in j} \left[\frac{n^{j}_{\mathrm{s}}}{N^{j}}\mathcal{S}^{j}_{i} + \left(1- \frac{n^{j}_{\mathrm{s}}}{N^{j}}\right)\mathcal{B}^{j}_{i} \right],
\end{equation}

\noindent where $i \in j$ indicates that the $i^{\mathrm{th}}$ event belongs to the $j^{\mathrm{th}}$ sample (IC-40, IC-59, IC-79 or IC-86I), $N^j$ is the total number of events in the $j^{\mathrm{th}}$ sample, $\mathcal{S}^j_i$ and $\mathcal{B}^{j}_i$ is the signal and background probability density function (PDF) respectively, $\gamma$ is the spectral index of the differential spectrum $dN/dE \propto E^{-\gamma}$ and is assumed to be the same for all the data samples. The total number of signal events $n_{\mathrm{s}}$ is the sum of the signal events coming from each sample $n^{j}_{\mathrm{s}}$. The fractional contribution of each $n_{\mathrm{s}}^j$ to the total $n_{\mathrm{s}}$ is fixed by the relative neutrino effective areas of each of the configurations (determined by detector simulation), and varies depending on the spectral index and the declination. The likelihood function in equation~(\ref{eq:likelihood}) is a function of the global common parameters $\gamma$ and $n_{\mathrm{s}}$ and the parameters specific for each time dependent search.

The signal PDF $\mathcal{S}^j_i$ is given by:

\begin{equation}
\label{eq:signalpdf}
\mathcal{S}^j_i = P^{\mathrm{signal},j}_{i} (|\vec{x}_i-\vec{x}_{\mathrm{s}}|,\sigma_{i})\cdot \mathcal{E}^{\mathrm{signal},j}_{i} (E_i, \delta_i, \gamma) \cdot \mathcal{T}^{\mathrm{signal},j}_{i},
\end{equation}

\noindent where $\vec{x}_{\mathrm{s}}$ is the source direction, $\vec{x}_i$ is the reconstructed direction of the event, $\sigma_{i}$ is the angular error estimate for the reconstruction, and $E_i$ and $\delta_i$ are the reconstructed energy and declination respectively. The term $P^{\mathrm{signal},j}_{i}$ is the spatial PDF (the point spread function) and $\mathcal{E}^{\mathrm{signal},j}_{i}$ is the energy PDF. These two terms are the same as for the IceCube time integrated searches and can be found in~\cite{juanan}. The time dependent signal PDF $\mathcal{T}^{\mathrm{signal},j}_{i}$ is specific for each of the different signal hypotheses and will be described in the corresponding sections.

Since the data are background dominated, the background PDF $\mathcal{B}^{j}_i$ can be estimated from the data itself and it has the form:

\begin{equation}
\label{eq:backgroundpdf}
\mathcal{B}^j_i = P^{\mathrm{bkg},j}_{i} (\delta_{i}) \cdot \mathcal{E}^{\mathrm{bkg},j}_{i} (E_i, \delta_i) \cdot \mathcal{T}^{\mathrm{bkg},j}_{i}.
\end{equation}

For sufficiently long time scales, the spatial PDF for the background $P_i^{\mathrm{bkg},j}(\delta_i)$ is uniform in right ascension as the Earth rotation averages over detector effects and follows the distribution of the data with declination. For short time scales (less than a day) the spatial PDF is no longer uniform in right ascension and shows a pattern when the neutrino directions align with the directions of geometrical symmetries of the non-uniform detector array. The effect disappears as the time scale grows large enough to allow the Earth rotation to average out the differences. The energy density function $\mathcal{E}^{\mathrm{bkg},j}_{i}$ can be found in~\cite{juanan}. The time probability $\mathcal{T}^{\mathrm{bkg},j}_{i}$ is taken to be flat since it was demonstrated in~\cite{mike} that the seasonal modulations of background atmospheric muons and neutrinos are negligible compared to possible measurable signals.

For all the searches presented here the test statistic $TS$ is defined as the maximum likelihood ratio:

\begin{equation}
    TS = -2 \log\Big[\frac{\mathcal{L}(n_{\mathrm{s}}=0)}{\mathcal{L}(\hat{n}_{\mathrm{s}},\hat{\gamma}_{\mathrm{s}},...)}\Big],
    \label{eq:ts}
\end{equation}

\noindent where $\mathcal{L}(n_{\mathrm{s}}=0)$ is the likelihood of the background-only (null) hypothesis and $\mathcal{L}(\hat{n}_{\mathrm{s}},\hat{\gamma}_{\mathrm{s}},...)$ is the likelihood evaluated with the best-fit values for the signal-plus-background hypothesis. The parameters denoted with carets are the best fit values of $n_{\mathrm{s}}$ and $\gamma_{\mathrm{s}}$ which are present in all searches\footnote{The two parameters, $n_{\mathrm{s}}$ and $\gamma_{\mathrm{s}}$, are present in all searches but their best fit values $\hat{n}_{\mathrm{s}}$ and $\hat{\gamma}_{\mathrm{s}}$ can be different for the different searches.} while the parameters specific to each time dependent search (represented by the ellipsis above) will be discussed in each search section. Estimates for the values of these parameters can be derived from the maximization of the likelihood function defined in equation~(\ref{eq:likelihood}).

The test statistic $TS$ as defined in equation~(\ref{eq:ts}) for the background-only data is expected to follow a~$\chi^2$-distribution with number of degrees of freedom reflecting the number of fitted parameters. This behavior of the $TS$ can be used to estimate the pre-trial p-value as the probability of the $TS$ value from the ~$\chi^2$-distribution.

Because our samples are background dominated, in order to translate the pre-trial p-values into post-trial p-values the time scrambled IceCube data from the same period can be used to generate background samples. The time scrambling procedure assigns to the events a random time within the period while keeping all other event properties (energy, local reconstructed coordinates, etc.) unchanged. In this way many different samples describing the background are obtained. About one thousand of these sets were generated and for each of them a p-value was obtained. Then the distribution of these p-values for the time scrambled data were compared to the p-value obtained for real data. From this the probability is derived that a random fluctuation will result in deviation at least as significant as our result for real data and this probability is called the post-trial p-value. The obtained post-trial p-value is stable with respect to deviations of background-only $TS$ from the assumed $\chi^2$ distribution 
and accounts for 
trial factors related to looking at many positions at the sky.

In analyzing the results it is convenient to characterize events by their time integrated weight $w_i$, defined as the ratio of the signal and background PDFs from equation~(\ref{eq:signalpdf}) and~(\ref{eq:backgroundpdf}) without the time PDF term:

\begin{equation}
\label{eq:signalweight}
w_i=\frac{P^{\mathrm{signal},j}_{i} (|\vec{x}_i-\vec{x}_{\mathrm{s}}|,\sigma_{i})\cdot \mathcal{E}^{\mathrm{signal},j}_{i} (E_i, \delta_i, \gamma)}{P^{\mathrm{bkg},j}_{i} (\delta_{i}) \cdot \mathcal{E}^{\mathrm{bkg},j}_{i} (E_i, \delta_i)},
\end{equation}

\noindent which will be later used to visualize the result in a way helpful for understanding whether the significance comes from the spatial or the time clustering.

For the time dependent searches\footnote{Except for the ``Search for Periodic Neutrino Emission from Binary Systems'' in Section~\ref{sec:per} which benefits from the signal being accumulated over many periods of the binary system.} it is natural to express the discovery potentials, sensitivities, and upper limits in terms of fluence, defined for an $E^{-2}$ spectrum as:

\begin{equation}
\label{eq:fluence}
f =\int^{t_{\mathrm{max}}}_{t_{\mathrm{min}}} dt \int^{E_{\mathrm{max}}}_{E_{\mathrm{min}}}dE \times E\frac{dN}{dE} = \Delta t \int^{E_{\mathrm{max}}}_{E_{\mathrm{min}}}dE \times E\frac{\Phi_0}{E^2}=\Delta t \Phi_0 \int^{E_{\mathrm{max}}}_{E_{\mathrm{min}}}\frac{1}{E}dE
\end{equation}

\noindent where $\Phi_0$ is the normalization of an average flux during the emission. The emission duration $\Delta t$ is defined by the time PDF. The integration limits $E_{\mathrm{min}}$ and $E_{\mathrm{max}}$ are set to the 5\% and 95\% resp. energy percentile of the event sample for the given declination band.

\section{All-Sky Time Scan} 
\label{sec:All-Sky Time Scan}

This is the most generic time dependent search among the ones presented in this paper. It is optimized to look for neutrino emission from a point-like source with limited duration. Because it aims for one-time flares it does not benefit from adding multiple datasets together. Here the results for the IC-59, IC-79 and IC-86I IceCube data are presented separately. The results of a similar search for the IC-40 configuration can be found in~\cite{mike}.

\subsection{Method}
\label{untrig_method}

The ``All-Sky Time Scan'' is an untriggered search since it is performed only using the neutrino data itself. For this search the whole sky\footnote{Since for declinations above +85$^\circ$ and below -85$^\circ$ the off-source region is very small and the statistics for data scrambling is too limited, this region is excluded from the analysis.} was divided into a grid of 0.1$^\circ \times 0.1^\circ$, much finer than the angular resolution of the detector, and the likelihood was maximized at each grid point. The expression for the likelihood is given by equation~(\ref{eq:likelihood}) (with each detector configuration analyzed separately). In equation~(\ref{eq:signalpdf}) the signal PDF $\mathcal{T}^{\mathrm{signal}}_{i}$ was chosen of the following form:

\begin{equation}
    \mathcal{T}^{\mathrm{signal}}_{i}=\frac{1}{\sqrt{2\pi}\sigma_{\mathrm{T}}}exp\left(-\frac{(t_i-T_0)^2}{2\sigma^2_{\mathrm{T}}}\right),
    \label{eq:time_pdf_allSky}
\end{equation}

\noindent where $t_i$ is the arrival time of the $i^{\mathrm{th}}$ event, $T_0$ and $\sigma_{\mathrm{T}}$ are the mean and the width of a Gaussian time structure, respectively. The Gaussian function was used as a smooth, general parametrization of a limited duration increase in the emission of a source. The best fit values of these parameters were obtained maximizing the likelihood function described in Section~\ref{sec:llh}.

Since within the limited livetime of a given sample it is possible to accommodate a larger number of flares as their duration decreases, an undesired bias towards finding a short flare is introduced. This could also be interpreted as a hidden extra trial factor affecting short flares. To account for it the test statistic $TS$ from equation~(\ref{eq:ts}) was modified. As explained in~\cite{method2}, an additional marginalization term $T/\sqrt{2\pi}\sigma$ that penalizes short flares compared to long ones was introduced. The test statistic including this modification is thus:

\begin{equation}
    TS = -2 \log\Big[\frac{T}{\sqrt{2\pi}\hat{\sigma}_{\mathrm{T}}}\times\frac{\mathcal{L}(n_{\mathrm{s}}=0)}{\mathcal{L}(\hat{n}_{\mathrm{s}},\hat{\gamma}_{\mathrm{s}},\hat{\sigma}_{\mathrm{T}},\hat{T}_0)}\Big],
    \label{eq:ts_allSky}
\end{equation}

\noindent where $\hat{n}_{\mathrm{s}},\hat{\gamma}_{\mathrm{s}},\hat{\sigma}_{\mathrm{T}},\hat{T}_0$ are the best-fit values and $T$ is the total livetime of the data taking period (either IC-59, IC-79 or IC-86I). For details on the numerical maximization procedure see~\cite{mike}.

\begin{figure}[!h]
  \centering
  \subfigure[]{\includegraphics[width=2.9in]{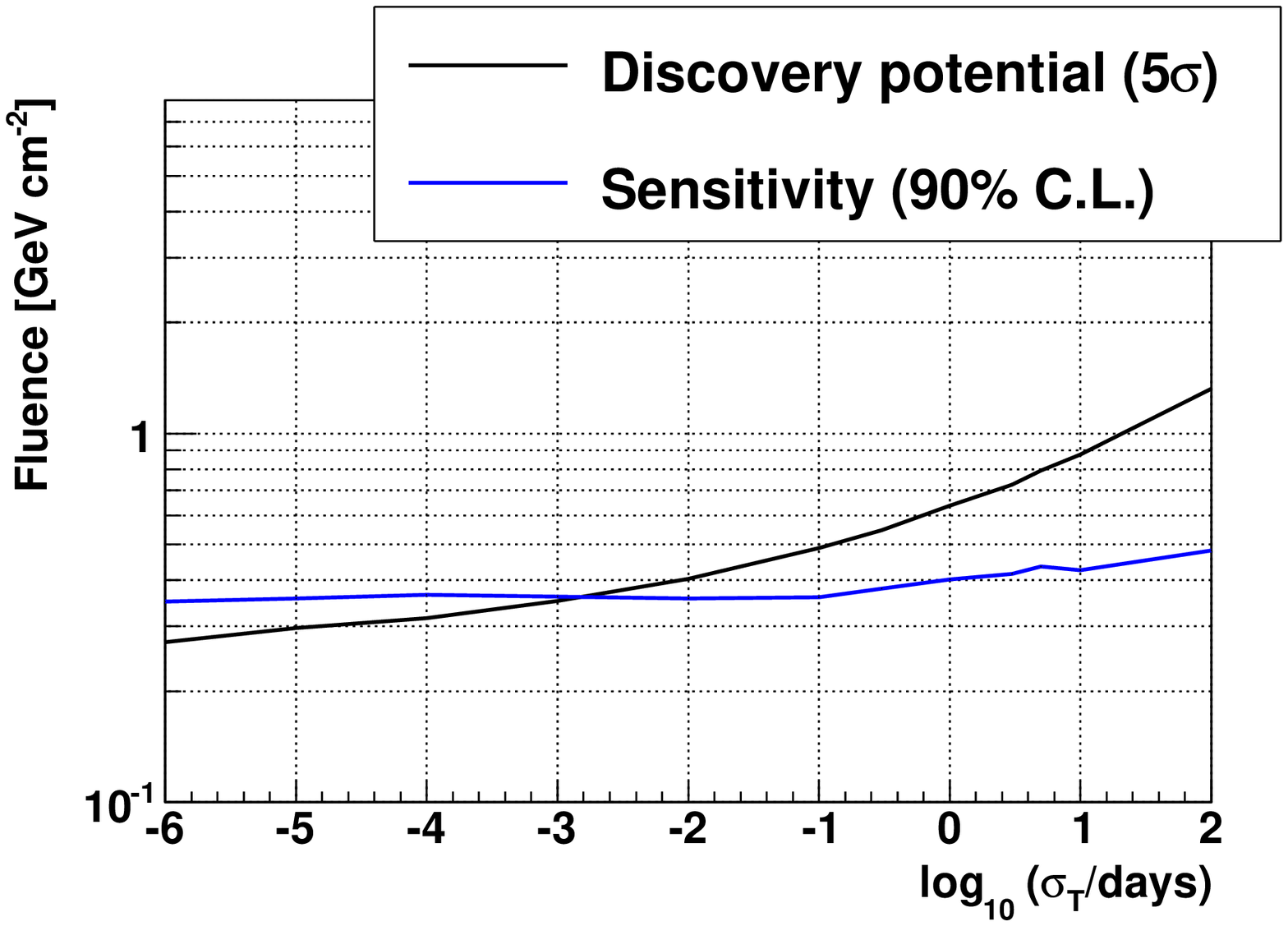}}
  \subfigure[]{\includegraphics[width=2.9in]{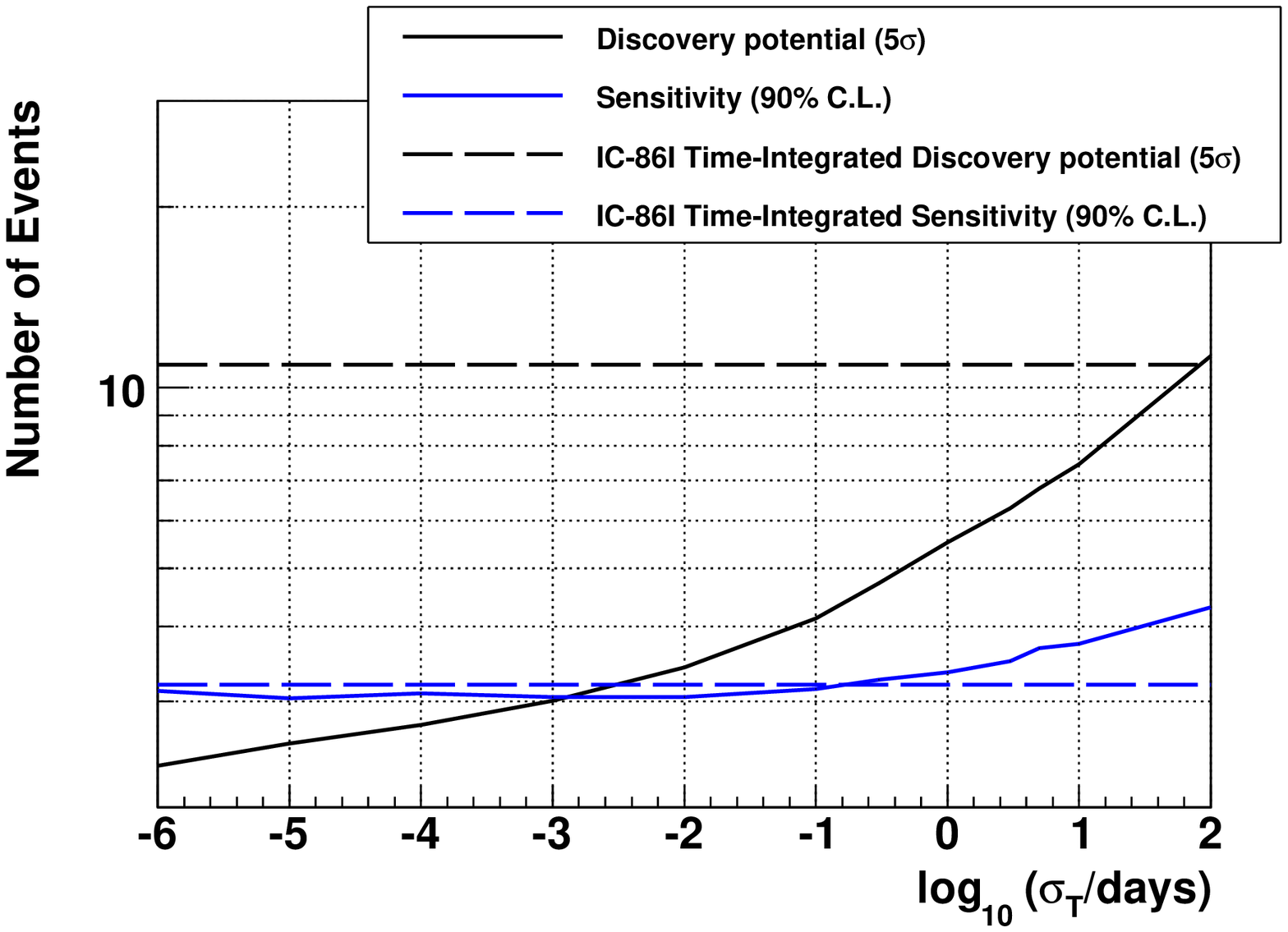}}
  \caption{The 5$\sigma$ discovery potential (signal required for 5$\sigma$ detection in 50\% of trials) and the sensitivity (90\% CL median upper limit) for IC-86I shown in terms of the fluence~(a) and the mean number of signal events~(b) for a fixed source at $+16^{o}$ declination (solid lines) with an $E^{-2}$ spectrum. The corresponding lines for the time integrated search are also shown. The time dependent search improves over the time integrated for flaring sources when solid lines become lower than dashed ones.}
  \label{fig:untriggered_discPot}
\end{figure}

The expected performance of this approach in terms of discovery potential and sensitivity is shown in Figure~\ref{fig:untriggered_discPot}. The figure compares this time dependent search with the standard time integrated all sky search for given declination. The ``All-Sky Time Scan'' search is better for flares with $\sigma_{\mathrm{T}}$ shorter than hundred days in terms of discovery potential and for flares with $\sigma_{\mathrm{T}}$ shorter than 6 hours in terms of sensitivity. As the flare duration $\sigma_{\mathrm{T}}$ gets shorter, the sensitivity levels out at around three events. For the calculation of the discovery potential at least two events are required in order to identify a flare. To generate a sample according to a Poissonian distribution with 50\% of the cases having two or more events, the Poissonian mean has to be equal to 1.68. Therefore for the shortest timescales the discovery potential will asymptotically approach this value causing it to drop below the sensitivity.

\subsection{Results}

\label{untrig_rez}
The maximization of the likelihood at each grid point in the sky results in a map of $TS$ values or, equivalently, a map of pre-trial p-values that serve as an estimate of the local significance of the best-fit parameters.  To address the question of whether any excess in the sky is significant, a correction for the trial factor involved in searching the entire sky had to be used. This was done as described at the end of Section~\ref{sec:llh} by repeating the analysis on time-scrambled data.

\subsubsection{IC-59 Results}

\begin{figure}[h]
\centering
\includegraphics[width=\textwidth]{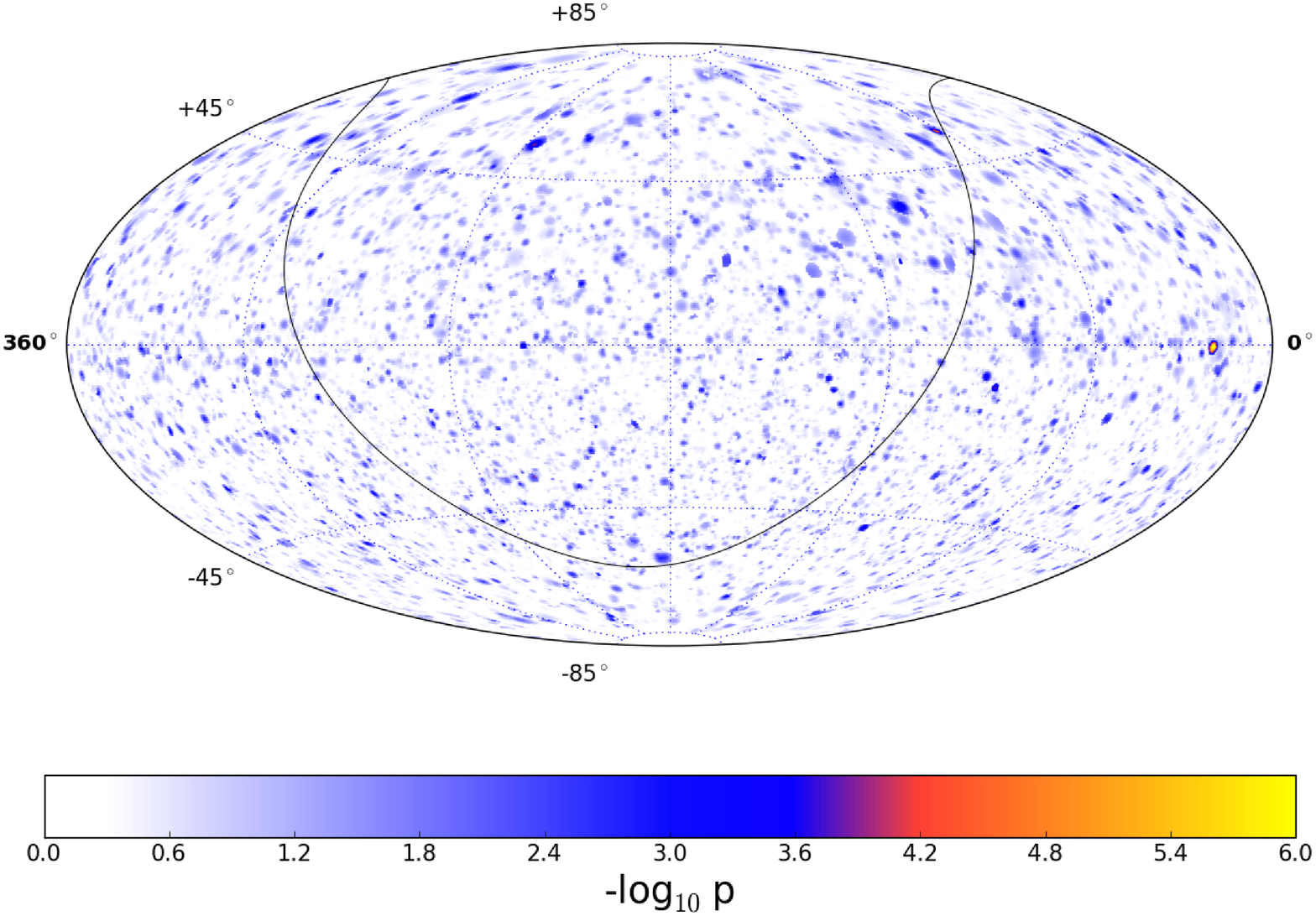}
\caption{IC-59 skymap in equatorial coordinates showing the pre-trial p-value for the best-fit flare hypothesis tested in each direction of the sky. The strongest Gaussian-like signal was found at (RA,~Dec.)~=~(21.35$^{\circ}$,~-0.25$^{\circ}$), with post-trial significance of p=1.4\%, see Table~\ref{tab:all_sky_rez} for details. The black curve is the galactic plane.}
\label{fig59_p}
\vspace{2cm}
\centering
\includegraphics[width=5.0in]{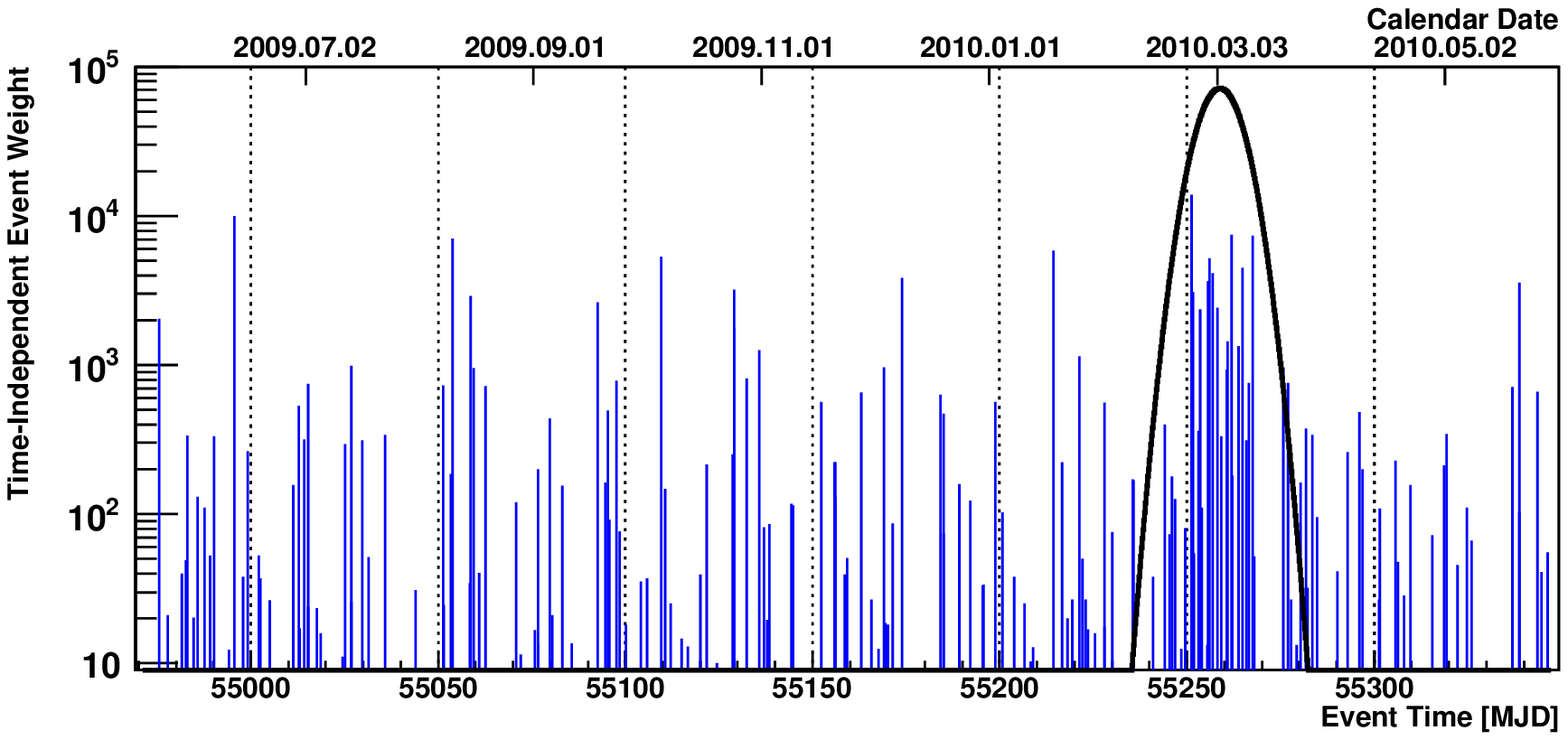}
\caption{The time integrated event weights $w_i$, defined in equation~(\ref{eq:signalweight}), evaluated for the IC-59 data at the location of the most significant flare. The best-fit Gaussian time PDF is shown in black (arbitrary scaling). See Table~\ref{tab:all_sky_rez} for details.}
\label{fig59_flareevts}
\end{figure}

Figure~\ref{fig59_p} shows the IC-59 skymap of pre-trial p-values for the all-sky search. The most significant point in the IC-59 data was found at (RA,~Dec.)~=~(21.35$^{\circ}$, -0.25$^{\circ}$). The peak occurred on MJD 55259 (2010 March 4), and had a width parameter $\hat{\sigma}_{\mathrm{T}}$ of 5.5 days, a soft spectral index of $\hat{\gamma}=3.9$, and 14.5 fitted signal events. The pre-trial p-value was $2.04\times10^{-7}$; a value at least as significant as this was found somewhere in the sky in 14 out of 1000 scrambled maps. Thus the post-trial p-value was 1.4\%, which was low but not significant evidence of an actual flare. When this analysis was repeated on the data for the following years (these results are presented in the following sections) the IC-59 hot spot significance decreased and it was not seen with high significance any more. Figure~\ref{fig59_flareevts} shows the time integrated event weights~$w_i$ at the position of maximum significance plotted throughout the year, a clustering near the time of the best-fit $\hat{T}_0$ is clearly visible. 
When a bin of radius $2^{\circ}$ and 13 days in time (the FWHM of the Gaussian) centered on the peak is considered, 13 events are found compared to an expected background of 1.7. 

\newgeometry{top=2cm}
\restoregeometry

%

\FloatBarrier
\subsubsection{IC-79 Results}

\begin{figure}[h]
  \centering
  \includegraphics[width=\textwidth]{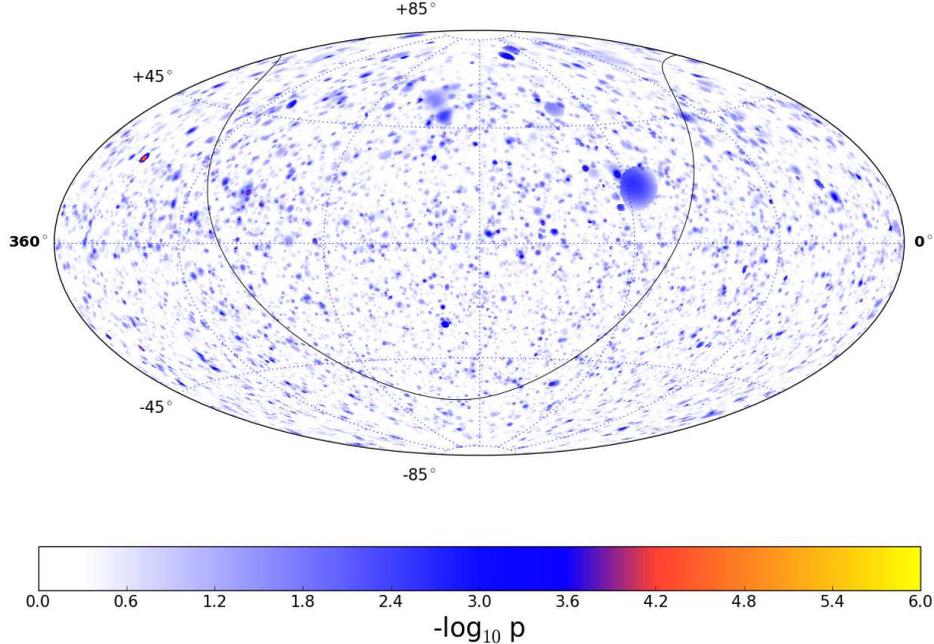}
  \caption{IC-79 skymap in equatorial coordinates showing the pre-trial p-value for the best-fit flare hypothesis tested in each direction of the sky. The strongest Gaussian-like signal was found at (RA,~Dec.)~=~(343.45$^{\circ}$,~-31.65$^{\circ}$), with post-trial significance of p=66\%, see Table~\ref{tab:all_sky_rez} for details. The black curve is the galactic plane.}
  \label{fig:ic79_untriggered_skyMap}
\end{figure}

Figure~\ref{fig:ic79_untriggered_skyMap} shows the IC-79 skymap of pre-trial p-values. The most significant deviation from the background-only hypothesis was found at (RA, Dec.)=(343.45$^{\circ}$, -31.65$^{\circ}$). The mean of the fitted Gaussian flare was at MJD~55466 (2010 September 27) with a width parameter $\hat{\sigma}_{\mathrm{T}}$ of 1.8 days, a soft spectral index of $\hat{\gamma}=3.95$, and 7.2 fitted signal events. The large blue spot in the upper right quadrant in Figure~\ref{fig:ic79_untriggered_skyMap} was caused by two events arriving very close in time, the consequence of which was an increased significance over a wider area to which those two events contribute.

The pre-trial p-value obtained for the IC-79 hotspot was $1.07\times10^{-5}$. The post-trial p-value was 66\%, consistent with a background-only hypothesis.

\begin{figure}[h] 
  \centering
  \includegraphics[width=5.in]{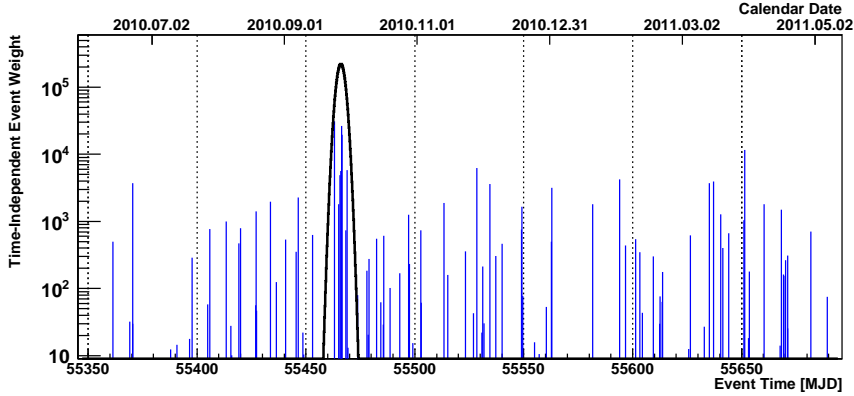}
  \caption{The time integrated event weights $w$, defined in equation~(\ref{eq:signalweight}), evaluated for the IC-79 data at the location of the most significant flare. The best-fit Gaussian time PDF is shown in black (arbitrary scaling). See Table~\ref{tab:all_sky_rez} for details.}
  \label{fig:ic79_untriggered_selectedEv}
\end{figure}

Figure~\ref{fig:ic79_untriggered_selectedEv} shows the time integrated event weights at the position of maximum significance plotted throughout the year with a hint of clustering recognizable at the fitted time.


\FloatBarrier
\subsubsection{IC-86I Results}

\begin{figure}[h]
\centering
\includegraphics[width=\textwidth]{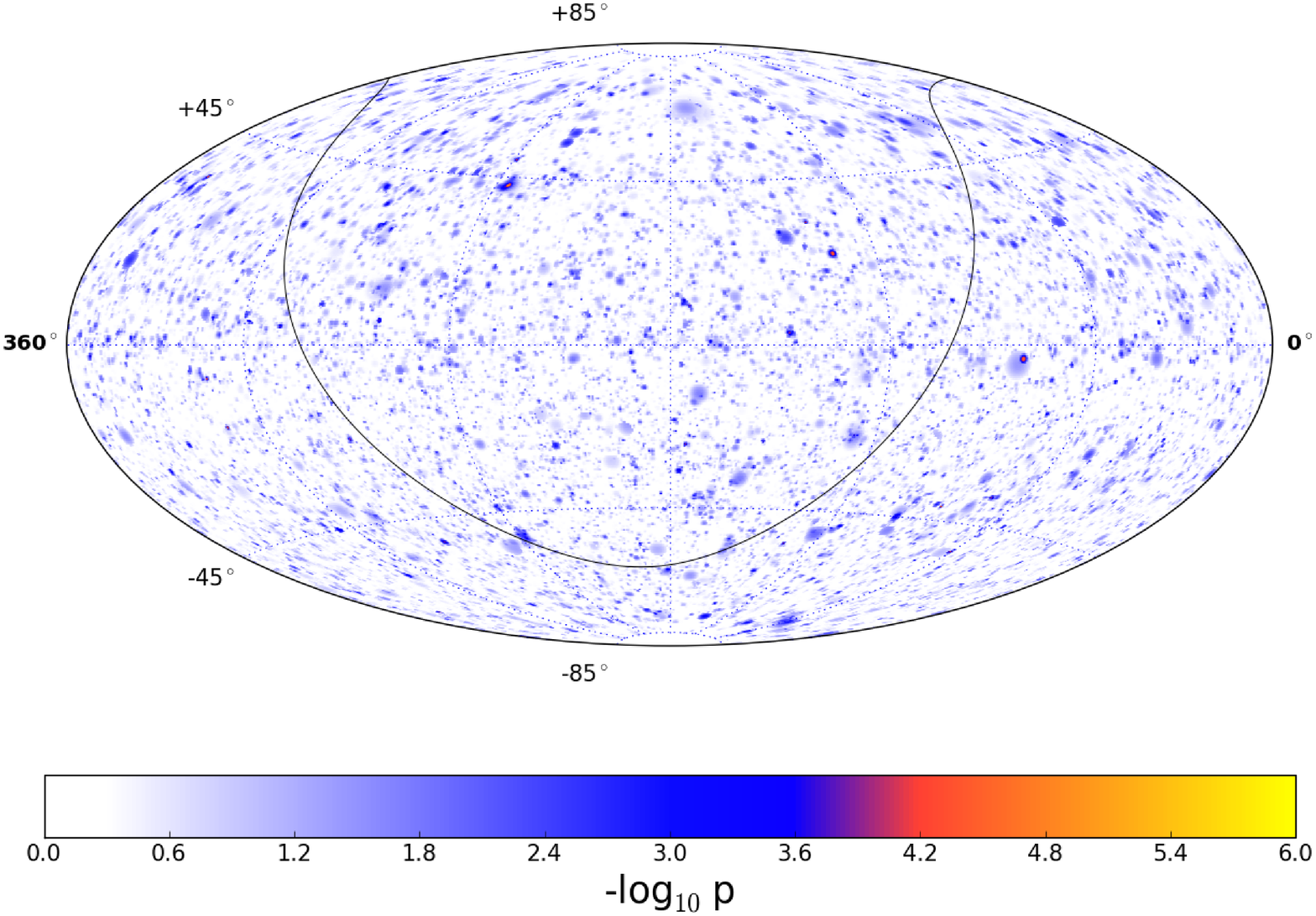}
\caption{IC-86I skymap in equatorial coordinates showing the pre-trial p-value for the best-fit flare hypothesis tested in each direction of the sky. The strongest Gaussian-like signal was found at (RA,~Dec.)~=~(235.95$^{\circ}$,~42.95$^{\circ}$), with post-trial significance of p=63\%, see Table~\ref{tab:all_sky_rez} for details. The black curve is the galactic plane.}
  \label{fig:ic86I_untriggered_skyMap}
\vspace{2cm}
\includegraphics[width=5.0in]{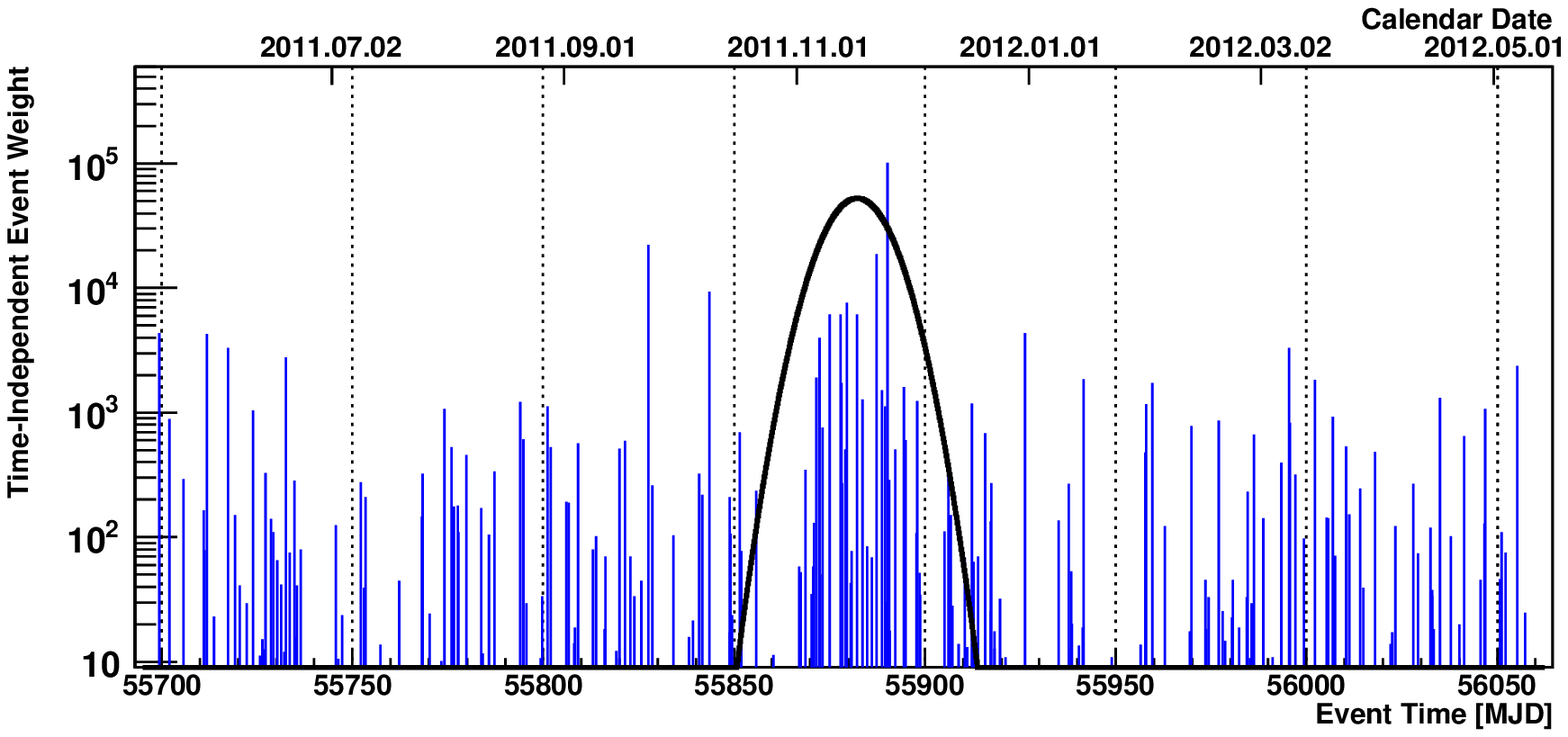}
\caption{The time integrated event weights $w$, defined in equation~(\ref{eq:signalweight}), evaluated for the IC-86I data at the location of the most significant flare. The best-fit Gaussian time PDF is shown in black (arbitrary scaling). See Table~\ref{tab:all_sky_rez} for details.}
  \label{fig:ic86I_untriggered_selectedEv}
\end{figure}

Figure~\ref{fig:ic86I_untriggered_skyMap} shows the skymap of the pre-trail p-values obtained for the IC-86I data, the most significant point was at (RA, Dec.)=(235.95$^{\circ}$, 42.95$^{\circ}$), and the fitted Gaussian parameters are MJD 55882 for the mean and 7.57 days for the width parameter $\hat{\sigma}_{\mathrm{T}}$. A spectral index of $\hat{\gamma}=2.0$, and 13.1 signal events were fitted. A pre-trial p-value of $1.06\times10^{-5}$ translates into a post-trial p-value of 63\%. The time integrated event weights for the region around the most significant point through the year are shown in Figure~\ref{fig:ic86I_untriggered_selectedEv}. 

\begin{table}[!h]
\centering
\vspace{1cm}
\small{
\begin{tabular}{lrrrrrrrr}
\toprule
\head{1.0cm}{Dataset} &\head{0.9cm}{RA} & \head{0.2cm}{Dec.} & \head{0.2cm}{$\hat{n}_{\mathrm{s}}$} & \head{0.5cm}{$\hat{\gamma}$} & \head{0.7cm}{$\hat{T}_0$ [MJD]}& \head{0.7cm}{$\hat{\sigma}_{\mathrm{T}}$ [days]} & \head{1.4cm}{p-value pre-trial} & \head{1.6cm}{p-value post-trial}\\ 
\midrule
IC-59 & 21.35$^{\circ}$  &  -0.25$^{\circ}$ & 14.5 & 3.9 & 55259 & 5.5 & $2.04\times10^{-7}$ & 1.4\%\\ 
IC-79 & 343.45$^{\circ}$ & -31.65$^{\circ}$ & 7.2 & 3.95 & 55466 & 1.8 & $1.07\times10^{-5}$ & 66.0\%\\
IC-86I& 235.95$^{\circ}$ &  42.95$^{\circ}$ & 13.1 & 2.0 & 55882 & 7.57& $1.06\times10^{-5}$ & 63.0\%\\ 
\bottomrule
\end{tabular}}
\caption{Location and best-fit parameters (number of signal events, spectral index, mean time and width parameter of Gaussian fit) of the most significant point (smallest pre-trial p-value) found in the data sample and the post-trial p-value (i.e. final significance).}
\label{tab:all_sky_rez}
\end{table}

\newgeometry{top=2cm}
\restoregeometry

%

\FloatBarrier
\section{Search for Triggered Multi-Messenger Flares}
\label{sec:fflares}
This search targets a set of astronomical objects which were observed to be in a flaring state by Fermi LAT~\cite{bib:Fermiweb}
during the time period analyzed in this study. The tested hypothesis is that the neutrino emission follows the intensity of the photon emission. If the photon emission has a hadronic origin, then the accelerated hadrons (protons or nuclei) interact with matter and produce neutral and charged pions, which produce $\gamma$-rays and neutrinos, respectively, when they decay.

Since Fermi~LAT provides continuous monitoring and the data is publicly available it is possible to make use of the measured lightcurves. A continuation of the search made during the previous period of IceCube data-taking~\cite{mike,ic59-time} is presented here and it was extended including the whole IC-59, IC-79 and IC-86I combined data samples. Also a more advanced de-noising method was implemented, which is described below, to better reconstruct the Fermi-LAT lightcurves.

\subsection{Method}
\label{fflares_method}

This search was performed only for selected objects in the sky. The criteria for selecting these objects (FSRQs, BL Lacs, etc.) were based on the Fermi-LAT photometric measurements. The Fermi LAT monitored source list~\cite{Fmonitored} was taken as a starting point. The first step was to retrieve the raw lightcurves from the Fermi Public Release data, using the analysis tools made available by the Fermi-LAT Collaboration. For each source the \textit{Fermi Science Tools v9r31p1} package~\cite{bib:FermiTools} was used to select photons within~$2^\circ$ of the source and to calculate the exposure. Photon events with zenith angles greater than $105^\circ$ were excluded to avoid contamination due to the Earth's albedo. For each source lightcurves with one day binning were obtained. 

The denoised lightcurve was used as time dependent signal PDF, $\mathcal{T}^{\mathrm{signal},j}_{i}$. The ``Bayesian Blocks'' method~\cite{bib:BBlocks,bib:BBlocks1} was applied to de-noise the lightcurves, implemented for the purpose of this analysis in the version described in~\cite{bib:BBlocks}. A simplified explanation of the method is that it splits the time axis of the lightcurve into blocks for which the flux variations are assumed to be compatible with Poisson distributions with a constant mean within each block. The criterion for deciding whether to split a section into two blocks or not is based on comparing how well the variations follow a single Poisson distribution or two Poisson distributions with different means.

\begin{figure}[!h]
  \centering
  \subfigure[$F_{\mathrm{B}}$=0.5]{\includegraphics[width=2.65in]{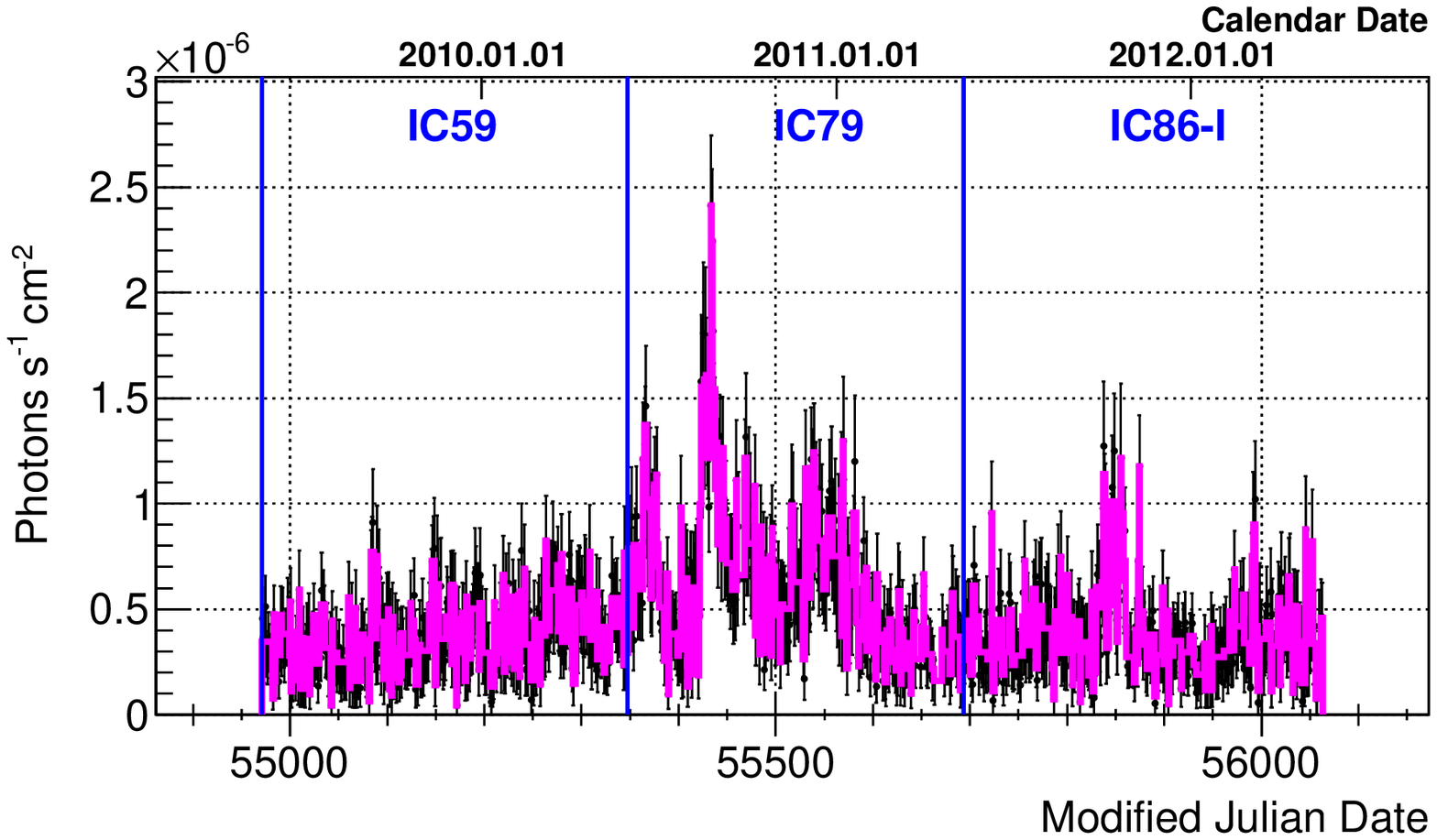}}
  \subfigure[$F_{\mathrm{B}}$=50.]{\includegraphics[width=2.65in]{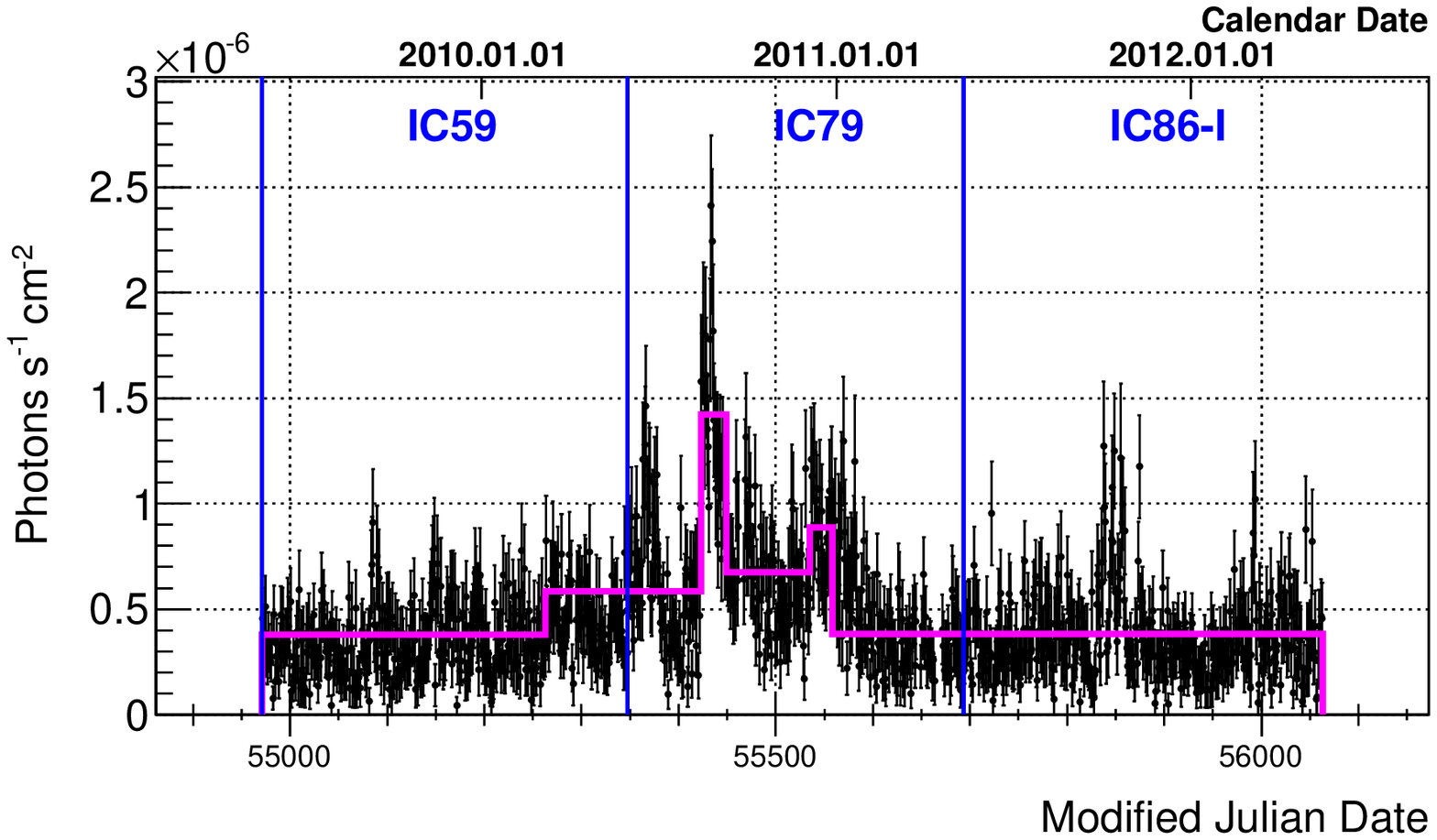}}
  \caption{The Bayesian Blocks method applied to the Fermi-LAT lightcurve of Ton 599 for two different values of the parameter $F_{\mathrm{B}}$ outside of the optimal range: in the left plot too small a value of $F_{\mathrm{B}}$ makes the denoised lightcurve (magenta line) follow all the statistical fluctuations of the lightcurve, while in the right plot too large a value of the parameter causes only six blocks to be identified.}
  \label{fig:BBlocksWrong}
\end{figure}

In order to optimize the performance of the Bayesian Blocks method on IceCube data the optimal value of the method parameter $F_{\mathrm{B}}$ had to be found. This parameter affects the method behavior in the following way. If the log-likelihood of two Poisson distributions is larger than the log-likelihood value for a single Poisson distribution by at least $F_{\mathrm{B}}$ the split is made. Too small values of $F_{\mathrm{B}}$ will cause denoised lightcurves to follow almost every point in the lightcurve while for too large values of $F_{\mathrm{B}}$ the denoised result will ignore important structures of the lightcurve. In Figure~\ref{fig:BBlocksWrong} two examples of the denoised lightcurves are shown for $F_{\mathrm{B}}$ being outside of the optimal range. To determine the value of the parameter $F_{\mathrm{B}}$ to be used in this analysis, a series of tests were performed to evaluate the performance of the Bayesian Blocks method as function of $F_{\mathrm{B}}$.

Realistic Fermi-LAT lightcurves were simulated in order to optimize the $F_{\mathrm{B}}$ parameter, using Fermi-LAT exposure data and assuming constant flux levels with Poissonian fluctuations. Folding in the Poisson distributed flux with the exposure gave a model of the background fluctuations including the correct statistical errors. On top of this background Gaussian shaped flares with random mean in time were injected. The background level was set to $0.5 \times 10^{-6}\:\mathrm{photons\:cm^{-2} s^{-1}}$ and was not varied. Instead the properties of the Gaussian flares were varied, i.e. strength of the flare, the time and the duration. The amplitude values were taken to be either equal to the background level or twice the background level. The tested widths were two days, five days and ten days. In addition, combinations of two injected flares were also tested.

\begin{figure}[!h]
  \centering
  \includegraphics[width=4.in]{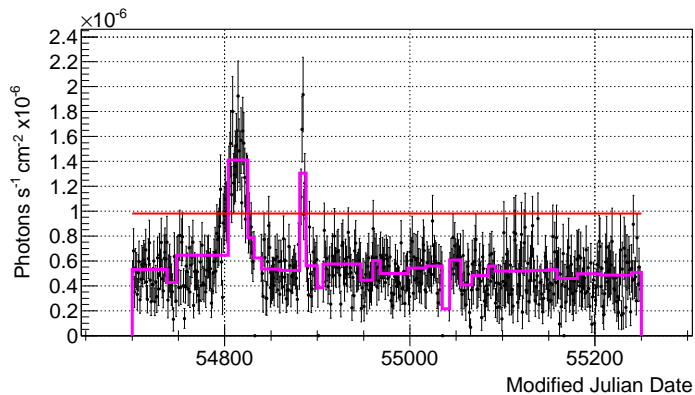}
  \caption{Example of the simulated background with two injected flares. In this example the noise level is $0.5\times10^{-6}\:\mathrm{photons\:cm^{-2} s^{-1}}$ and the red horizontal line indicates a three standard deviations upward fluctuation of the background. Two flares were injected on top of the background with widths of two and ten days with amplitudes of $10^{-6}\:\mathrm{photons\:cm^{-2} s^{-1}}$. The solid magenta line is the result of the Bayesian Blocks denoising procedure with the value of $F_{\mathrm{B}}=5.0$. Both flares are clearly visible and both were successfully recognized by the method.}
  \label{fig:sim2flares}
\end{figure}

For each of the tested combinations of a flare with some amplitude and duration, hundreds of random instances of lightcurves with injected flares and background were produced. For each of these the lightcurve was denoised, scanning over different values of $F_{\mathrm{B}}$, and two quantities were calculated as a function of $F_{\mathrm{B}}$: the rate of finding a fake flare, i.e. a flare at a position where none was injected; and the rate of finding the injected flare. For this purpose ``successful flare finding'' was defined as the denoised lightcurve reaching above a three standard deviations upward fluctuation of the background. In Figure~\ref{fig:sim2flares} an example is shown of two flares of different duration being injected and successfully recognized by the Bayesian Blocks method.

\begin{figure}[!h]
  \centering
  \includegraphics[width=4.in]{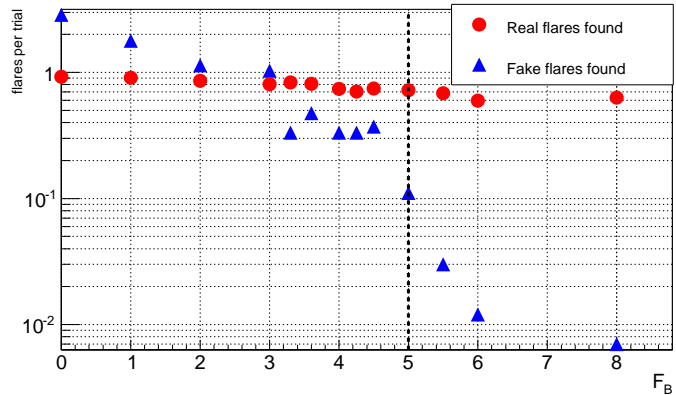}
  \caption{Example of the performance of the Bayesian Blocks method as a function of the $F_{\mathrm{B}}$ parameter for a single flare injected with width of two days and amplitude of $10^{-6}\:\mathrm{photons\:cm^{-2} s^{-1}}$. The blue squares indicate the rate (number flares per trial) at which fake flares are found. In red circles the rate for finding the injected flare is shown. The value of $F_{\mathrm{B}}$ chosen to be used for the analysis is 5.0, shown as vertical dashed line in the plot.}
  \label{fig:FBscan}
\end{figure}

\begin{figure}[!h]
  \centering
  \includegraphics[width=4.in]{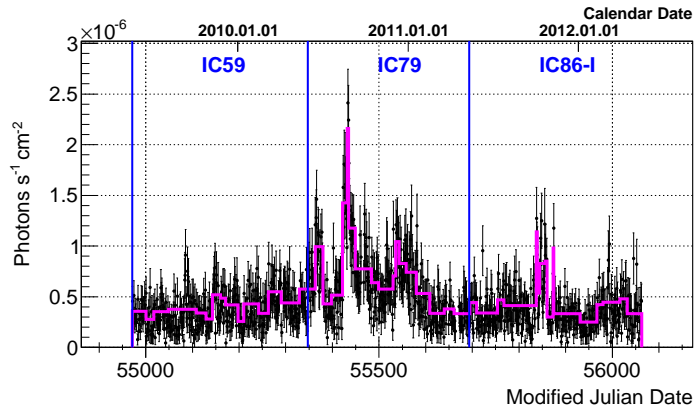}
  \caption{Example of a denoised lightcurve (solid line) together with the original data (black data points) for blazar Ton~599.}
  \label{fig:BBlocksRight}
\end{figure}

After evaluating the $F_{\mathrm{B}}$ scans for the different simulated combinations of background and injected flares the value of $F_{\mathrm{B}}$ to be used in this analysis was set to 5. Using this value the rate of fake flares drops significantly while the rate of finding the injected flares is still high. Figure~\ref{fig:FBscan} shows an example of the $F_{\mathrm{B}}$ scan and in Figure~\ref{fig:BBlocksRight} an example of the denoising method applied to a real lightcurve for one of the candidate sources with the chosen value of $F_{\mathrm{B}}=5$ is shown.

At this point some selection criteria were applied on the Fermi LAT monitored list of sources, with the aim of selecting lightcurves with flaring behavior and significant enhancement of the flux over the average level. The first criterion was that the denoised lightcurve flux must reach above $10^{-6}\:\mathrm{photons\:cm^{-2} s^{-1}}$ during IC-86I\footnote{The requirement was for the flux to reach above $10^{-6}\:\mathrm{photons\:cm^{-2} s^{-1}}$ specifically during IC-86I and not during the whole analyzed time window (IC-59, IC-79 and IC-86I together) because the IC-59 and IC-79 data were analyzed before, finding no significant results~\cite{ic59-time}.}. The second selection criterion aims at identifying denoised lightcurves exhibiting significant variations in time. In order to quantify this an 11-day running mean\footnote{The value of 11-day for the running mean comes from the fact that the denoised lightcurve has one day binning. Thus for calculating the mean we take the value for 
the 
current bin and five bins before and after for 11-day in total.} was calculated for each of the candidate sources. The maximum spread of the running mean was then divided by the mean over the entire 3-year data taking period, resulting in a measure of the maximum relative time variation. Only sources for which this maximum relative time variation was greater than 0.5 were selected.

After selecting a set of potentially interesting neutrino sources the statistical approach described in Section~\ref{sec:llh} was applied. The general form of the likelihood function given by equation~(\ref{eq:likelihood}) was used and the signal PDF as defined in equation~(\ref{eq:signalpdf}) was obtained using the denoised Fermi~LAT lightcurves. For each candidate source the likelihood function was maximized with respect to the number of signal events $n_{\mathrm{s}}$, the power law index $\gamma$, the time lag $D_{\mathrm{t}}$ and flux threshold $f_{\mathrm{th}}$. The time lag parameter allows for a time offset between the photon lightcurve and the neutrino PDF of up to $\pm$0.5~days. Since neutrinos are expected to be produced in the high activity states of the sources the flux threshold was varied during the maximization procedure. Once the threshold changes, the time PDF is redefined, setting it equal to zero below the threshold and normalizing to unity what was left above the threshold. The test statistic for this search is given by the 
maximum likelihood ratio:

\begin{equation}
    TS = -2 \log\Big[\frac{\mathcal{L}(n_{\mathrm{s}}=0)}{\mathcal{L}(\hat{n}_{\mathrm{s}},\hat{\gamma}_{\mathrm{s}},\hat{D}_{\mathrm{t}},\hat{f}_{\mathrm{th}})}\Big],
    \label{eq:ts_bflares}
\end{equation}

\noindent where $\hat{n}_{\mathrm{s}},\hat{\gamma}_{\mathrm{s}},\hat{D}_{\mathrm{t}},\hat{f}_{\mathrm{th}}$ are the best-fit values for the number of signal events, the power law index, the time lag and time PDF threshold.

\subsection{Results}
\label{fflares_rez}

Using the selection criteria above, the list of sources in Table~\ref{tab:bflrares} was selected. The most significant deviation from the background-only hypothesis was observed for the Quasar PKS 2142-75 at (RA, Dec.)=(326.8$^{\circ}$, -75.6$^{\circ}$). To evaluate the post-trial p-value time-scrambled samples of the IceCube events were generated and the analysis was repeated on them for all the selected sources. Then the most significant result for each of the scrambled datasets was identified and its significance compared to the p-value for our most significant IceCube result. In 77\% of the scrambled sets the p-value was equal to or smaller than the most significant p-value observed in the non-scrambled data and therefore it is well compatible with background fluctuations.

\begin{figure}[!h]
\centering
  \includegraphics[width=5.in]{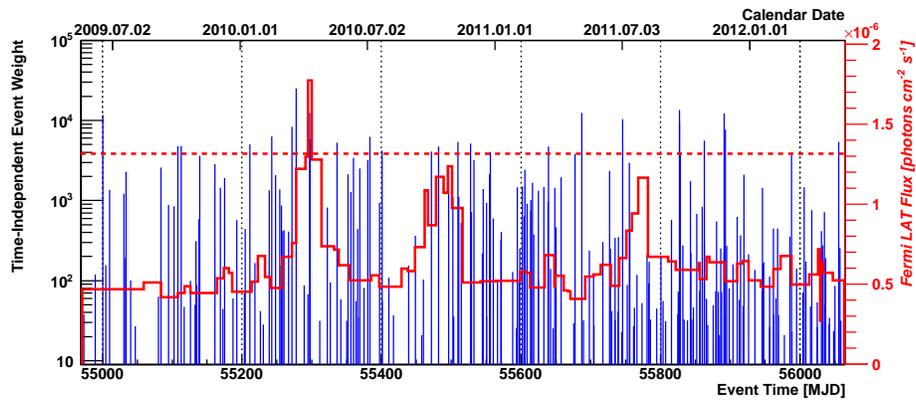}
  \caption{The denoised Fermi LAT lightcurve for PKS 2142-75 is shown with the red line and the red dashed horizontal line indicates the fit result for the flux threshold. For the lightcurve and the flux threshold the red scale on the right is used on the y-axis. The blue vertical lines are drawn at the times of measured IceCube events and the height indicates the event weights $w_i$ defined in Eg.~\ref{eq:signalweight} on the left scale. Only events in the periods when the lightcurve is above the best-fit flux threshold contribute to the significance.}
  \label{fig:SelectedSignalRegPKS2142}
\end{figure}

Figure~\ref{fig:SelectedSignalRegPKS2142} shows, for PKS 2142-75, the best-fit flux threshold together with the denoised lightcurve and the IceCube event weights $w_i$ defined in equation~(\ref{eq:signalweight}). In this figure one can see that the fit prefers a high flux threshold value, therefore reducing the time PDF to be non-zero only in a narrow time interval, leading to a low best-fit signal strength of $\hat{n}_{\mathrm{s}}=1.9$

\clearpage
\pagestyle{empty}
\newgeometry{top=1cm}
\voffset=1cm
\LTcapwidth=22cm
\begin{landscape}
\centering
\begin{longtable}{|l|r|r|r|r|r|r|r|r|r|r|r|}
\caption{Results of the triggered multi-messenger flare search. Besides the names and the equatorial coordinates of the sources the best-fit values for the likelihood function parameters $\hat{n}_{\mathrm{s}},\hat{\gamma}_{\mathrm{s}},\hat{D}_{\mathrm{t}},\hat{f}_{\mathrm{th}}$ are given. The p-values are pre-trial. The columns Duration (i.e.~the total amount of time for which the Fermi~LAT lightcurve is above the threshold $\hat{f}_{\mathrm{th}}$),~$\log( E_{\mathrm{min}})$, and $\log (E_{\mathrm{max}}$) list the values used for calculating the 90\%~C.~L. fluence upper limit. The fluence upper limits were calculated with time dependence corresponding to the fitted signal in the likelihood (i.e. non-zero only when the Fermi~LAT lightcurve is above the threshold $\hat{f}_{\mathrm{th}}$), and are hence provided only for sources that were positive fluctuations. These limits were calculated for a flux with spectral index of 2 irrespective of the actual best fit value. For sources with under-fluctuation, the number of fitted signal events $\hat{n}_{\mathrm{s}}$ is zero.}
\label{tab:bflrares}\\
\endfirsthead
\multicolumn{10}{c}%
{{\bfseries \tablename\ \thetable{} -- continued from previous page}} \\
\hline
\head{.6cm}{\multirow{ 2}{*}{\tabfont Name}} & \head{1.0cm}{\tabfont Dec. [$^{\circ}$]} & \head{.6cm}{\tabfont RA [$^{\circ}$]} & \head{1.15cm}{\multirow{ 2}{*}{\tabfont p-value}} & \head{1.0cm}{\multirow{ 2}{*}{\tabfont $\hat{n}_{\mathrm{s}}$}} & \head{1.0cm}{\multirow{ 2}{*}{\tabfont $\hat{\gamma}_{\mathrm{s}}$}} & \head{1.0cm}{\tabfont $\hat{D}_{\mathrm{t}}$ [days]} & \head{1.8cm}{\tabfont $\hat{f}_{\mathrm{th}}$ [$\mathrm{ph\:cm^{-2} s^{-1}}$]} & \head{1.2cm}{\tabfont Duration [days]} & \head{1.7cm}{\multirow{ 2}{*}{\tabfont $\log\left(\dfrac{E_{\mathrm{min}}}{\mathrm{GeV}}\right)$}} & \head{1.7cm}{\multirow{ 2}{*}{\tabfont $\log\left(\dfrac{E_{\mathrm{max}}}{\mathrm{GeV}}\right)$}} & \head{2.4cm}{\tabfont Fluence Limit\newline [GeV cm$^{-2}$]}\\
\hline \hline
\endhead
\hline \multicolumn{12}{|c|}{{Continued on next page}} \\ \hline
\endfoot
\hline
\endlastfoot
\toprule
\head{.6cm}{\multirow{ 2}{*}{\tabfont Name}} & \head{1.0cm}{\tabfont Dec. [$^{\circ}$]} & \head{.6cm}{\tabfont RA [$^{\circ}$]} & \head{1.15cm}{\multirow{ 2}{*}{\tabfont p-value}} & \head{1.0cm}{\multirow{ 2}{*}{\tabfont $\hat{n}_{\mathrm{s}}$}} & \head{1.0cm}{\multirow{ 2}{*}{\tabfont $\hat{\gamma}_{\mathrm{s}}$}} & \head{1.0cm}{\tabfont $\hat{D}_{\mathrm{t}}$ [days]} & \head{1.8cm}{\tabfont $\hat{f}_{\mathrm{th}}$ [$\mathrm{ph\:cm^{-2} s^{-1}}$]} & \head{1.2cm}{\tabfont Duration [days]} & \head{1.7cm}{\multirow{ 2}{*}{\tabfont $\log\left(\dfrac{E_{\mathrm{min}}}{\mathrm{GeV}}\right)$}} & \head{1.7cm}{\multirow{ 2}{*}{\tabfont $\log\left(\dfrac{E_{\mathrm{max}}}{\mathrm{GeV}}\right)$}} & \head{2.4cm}{\tabfont Fluence Limit\newline [GeV cm$^{-2}$]}\\
\midrule
PKS 2142-75    &   -75.6 &   326.8       &   0.02  &   1.9  &   3.95  &   -0.40  &   1.32e-06 &    5 & 5.0 & 7.8 & 8.10 \\
PKS 0235-618   &   -61.6 &    39.2       &   - & 0 & -&-&-&-&-&-&-\\
PMN J1038-5311 &   -53.2 &   159.7       &   0.47  &   0.97 &   2.25  &   0.50   &   0        & 1075 & 5.2 & 7.8 & 9.13 \\
Fermi J1717-5156 & -51.9 &   259.4       &   0.50  &   0.32 &   3.95  &   0.15   &   0        & 1075 & 5.2 & 7.8 & 8.38 \\
PKS 2326-502   &   -49.9 &   352.3       &   - & 0 & -&-&-&-&-&-&-\\
PKS 1424-41    &   -42.1 &   217.0       &   0.34  &   3.02 &   2.95  &   -0.37  &   8.15e-07 &  910 & 5.2 & 7.9 & 9.73 \\
PKS 0920-39    &   -40.0 &   140.7       &   0.10  &   6.38 &   3.25  &   -0.44  &   7.93e-07 &  684 & 5.1 & 7.8 & 11.60 \\
PKS 0426-380   &   -37.9 &    67.2       &   - & 0 & -&-&-&-&-&-&-\\
PKS 0402-362   &   -36.1 &    61.0       &   - & 0 & -&-&-&-&-&-&-\\
PMN J2250-2806 &   -28.1 &   342.7       &   - & 0 & -&-&-&-&-&-&-\\
PKS 2255-282   &   -28.0 &   344.5       &   - & 0 & -&-&-&-&-&-&-\\
PKS 1244-255   &   -25.8 &   191.7       &   0.08  &   0.98 &   3.22  &   0.03   &   1.73e-06 &    2 & 5.1 & 7.9 & 3.51 \\
PKS 1622-253   &   -25.5 &   246.4       &   0.49  &   0.54 &   2.75  &   -0.20  &   2.28e-06 &   82 & 5.1 & 7.9 & 2.37 \\
PMN J1626-2426 &   -24.4 &   246.8       &   0.30  &   3.28 &   2.15  &   -0.08  &   1.98e-06 &  269 & 5.2 & 7.8 & 3.58 \\
PKS 0454-234   &   -23.4 &    74.3       &   0.16  &   1.40 &   2.05  &   -0.30  &   2.38e-06 &    1 & 5.1 & 7.8 & 2.92 \\
PKS 1830-211   &   -21.1 &   278.4       &   - & 0 & -&-&-&-&-&-&-\\
PMN J2345-1555 &   -15.9 &   356.3       &   - & 0 & -&-&-&-&-&-&-\\
Fermi J1532-1321 & -13.4 &   233.2       &   0.45  &   2.56 &   2.85  &   0.40   &   0        & 1075 & 4.9 & 7.8 & 2.57 \\
PKS 1730-130   &   -13.1 &   263.3       &   - & 0 & -&-&-&-&-&-&-\\
PKS 0727-11    &   -11.7 &   112.6       &   0.09  &   7.74 &   3.55  &   -0.06  &   8.64e-07 & 1069 & 4.8 & 7.8 & 3.43 \\
PKS 1346-112   &   -11.5 &   207.4       &   0.18  &   1.37 &   2.21  &   -0.50  &   1.50e-06 &    2 & 4.8 & 7.8 & 1.49 \\
PKS 1510-089   &    -8.8 &   228.2       &   0.45  &   1.68 &   3.56  &   -0.17  &   0        & 1075 & 4.5 & 7.7 & 1.63 \\
3C 279         &    -5.8 &   194.0       &   - & 0 & -&-&-&-&-&-&-\\
PKS 2320-035   &    -3.3 &   350.9       &   0.12  &   2.67 &   2.50  &   0.28   &   1.34e-06 &    4 & 3.5 & 7.3 & 0.40 \\
PMN J0948+0022 &     0.4 &   147.2       &   0.27  &   7.12 &   3.95  &   0.50   &   0        & 1075 & 3.5 & 7.2 & 0.72 \\
3C 273         &     2.1 &   187.3       &   0.26  &   1.65 &   1.85  &   -0.36  &   1.12e-06 &  201 & 3.4 & 7.0 & 0.38 \\
PMN J0505+0416 &     4.3 &    76.4       &   - & 0 & -&-&-&-&-&-&-\\
J123939+044409 &     4.7 &   189.9       &   - & 0 & -&-&-&-&-&-&-\\
OG 050         &     7.5 &    83.2       &   - & 0 & -&-&-&-&-&-&-\\
CTA 102        &    11.7 &   338.2       &   0.26  &   3.58 &   3.95  &   0.10   &   7.14e-07 &  128 & 3.2 & 6.4 & 0.41 \\
3C 454.3       &    16.1 &   343.5       &   0.28  &   1.94 &   3.65  &   -0.45  &   1.99e-06 &   10 & 3.2 & 6.3 & 0.34 \\
OX 169         &    17.7 &   325.9       &   0.20  &   2.70 &   3.95  &   -0.50  &   9.45e-07 &   29 & 3.2 & 6.2 & 0.44 \\
OJ 287         &    20.1 &   133.7       &   - & 0 & -&-&-&-&-&-&-\\
PKS B1222+216  &    21.4 &   186.2       &   - & 0 & -&-&-&-&-&-&-\\
Crab Pulsar    &    22.0 &    83.6       &   - & 0 & -&-&-&-&-&-&-\\
4C 28.07       &    28.8 &    39.5       &   - & 0 & -&-&-&-&-&-&-\\
Ton 599        &    29.2 &   179.9       &   - & 0 & -&-&-&-&-&-&-\\
B2 1520+31     &    31.7 &   230.5       &   0.42  &   3.75 &   2.15  &   0.44   &   0        & 1075 & 3.0 & 5.8 & 0.78 \\
B2 1846+32     &    32.3 &   282.1       &   0.37  &   1.57 &   3.95  &   0.37   &   8.58e-07 &   81 & 3.0 & 5.8 & 0.43 \\
B2 0619+33     &    33.4 &    95.7       &   - & 0 & -&-&-&-&-&-&-\\
1H 0323+342    &    34.2 &    51.2       &   0.47  &   1.52 &   2.15  &   -0.25  &   5.40e-07 & 1051 & 3.0 & 5.8 & 0.71 \\
4C 38.41       &    38.1 &   248.8       &   0.23  &   11.55&   2.55  &   0.50   &   0        & 1075 & 3.0 & 5.7 & 1.19 \\
3C 345         &    39.7 &   250.4       &   0.09  &   4.0  &   2.18  &   0.16   &   1.07e-06 &   61 & 3.0 & 5.7 & 0.83 \\
NGC 1275       &    41.5 &    50.0       &   0.15  &   2.67 &   3.95  &   -0.40  &   1.25e-06 &   24 & 3.0 & 5.7 & 0.60 \\
BL Lac         &    42.3 &    330.7      &   - & 0 & -&-&-&-&-&-&-\\
B3 1343+451    &    44.9 &   206.4       &   0.30  &   0.82 &   1.55  &   0.19   &   1.29e-06 &   14 & 3.0 & 5.6 & 0.50 \\
4C 49.22       &    49.5 &   178.4       &   0.11  &   4.72 &   3.95  &   0.25   &   4.29e-07 &   99 & 3.0 & 5.6 & 0.66 \\
NRAO 676       &    50.8 &   330.4       &   - & 0 & -&-&-&-&-&-&-\\
BZU J0742+5444 &    54.7 &   115.7       &   - & 0 & -&-&-&-&-&-&-\\
S4 1849+67     &    67.1 &   282.3       &   0.25  &   11.70&   3.75  &   0.18   &   2.90e-07 & 1046 & 2.9 & 5.2 & 1.27 \\
S5 0836+71     &    70.9 &   130.4       &   0.23  &   3.13 &   3.95  &   -0.32  &   5.96e-07 &  109 & 2.9 & 5.2 & 0.75 \\
PKS 0716+714   &    71.4 &   110.4       &   - & 0 & -&-&-&-&-&-&-\\
S5 1803+78     &    78.5 &   270.2       &   0.19  &   3.98 &   3.95  &   0.12   &   7.92e-07 &   26 & 2.9 & 5.2 & 0.95 \\
\bottomrule
\end{longtable}
\end{landscape}
\restoregeometry
\clearpage
\pagestyle{plain}
\section{Search for Triggered Flares with Sporadic Coverage}
\label{sec:sflares}

In the searches presented in the previous section neutrino emission was assumed to follow the $\gamma$-ray photon flux of the source, using lightcurves extracted from Fermi-LAT data in the energy range from 100 MeV to 300 GeV as templates. A complementary search was performed to cover a potentially interesting group of sources that did not pass the selection criteria in Section~\ref{fflares_method}. These are sources which exhibited flares in the TeV range, but which did not show significant activity in the lower energy range covered by the Fermi-Lat lightcurves. For these flares Astronomer's~Telegrams (ATel) were issued by imaging air-Cherenkov telescopes such as H.E.S.S., MAGIC or VERITAS. As explained in Section~\ref{sec:sources}, such orphan flares, exhibiting TeV emission while lacking emissions in the lower energy region, may be interesting for hadronic 
models, if the lack of emission is real and not due to limited exposure of experimental observations~\cite{orphan1,orphan2,orphan3}. This search was performed for the IC-79 and IC-86I periods only since the data from previous periods have already been analyzed~\cite{mike}.

\subsection{Method}
\label{oflares_method}

Once more the general form of equation~(\ref{eq:likelihood}) was used for the likelihood function but here the time dependent part of the signal PDF was a normalized box function. The box shaped time PDF was defined as being equal to zero in the whole period except for the flare time reported by the ATel plus a one day margin before and after. The use of detailed lightcurves was impossible in these cases because there is no continuous monitoring of TeV observations, like Fermi-LAT provides at lower energies.

Since the duration of the flare was fixed, the likelihood function was maximized only with respect to the spectral index $\gamma$ and the number of signal events $n_{\mathrm{s}}$. 

\subsection{Results}

In Table~\ref{tab:table_box} the candidate sources are listed. These were selected from the reports found in the ATel indicated in the table. Two sources, 1ES 0806+524 during IC-79 and PG1553+113 during IC-86I, showed a positive fluctuation over the background but the p-values are compatible with the background.

\begin{table}[!h]
\begin{adjustwidth}{-1in}{-1in}
\centering
\begin{tabular}{|l|l|l|l|l|l|l|}
\toprule
\head{0.8cm}{Season} & \head{1.0cm}{Source} & \head{1.9cm}{ATel num.} & \head{3.6cm}{Time range in MJD} & \head{0.6cm}{$\hat{n}_{\mathrm{s}}$} & \head{0.6cm}{$\hat{\gamma}_{\mathrm{s}}$} & \head{1.3cm}{p-value} \\
\midrule
\multirow{3}{*}{IC-79} & 1ES 0806+524 & 3192 & 55615 \textless~T~\textless 55618 & 0.78 & 3.95 & 0.24 \\
 &HESS J0632+057 & 3153, 3161 & 55598 \textless~T~\textless 55602 & 0 & - & - \\
 &1ES 1215+303 & 3100 & 55562 \textless~T~\textless 55565 & 0 & - & - \\
\hline
\multirow{2}{*}{IC-86I} & PG1553+113 & 4069 & 56036 \textless~T~\textless 56039 & 0.8 & 3.95 & 0.23 \\
 & BL Lacertae & 3459 & 55739 \textless~T~\textless 55742 & 0 & - & - \\
\bottomrule
\end{tabular}
\end{adjustwidth}
\caption{Source candidates selected for the ``Search for Triggered Flares with Sporadic Coverage''. The best-fit values for the number of signal events $\hat{n}_{\mathrm{s}}$ and the spectral index $\hat{\gamma}_{\mathrm{s}}$ are listed. The p-values are pre-trial, the dashes indicate that the source best-fit was for zero signal events.}
\label{tab:table_box}
\end{table}
\FloatBarrier
\section{Search for Periodic Neutrino Emission from Binary Systems} 
\label{sec:per}

X-Ray binaries that have also been observed to emit TeV $\gamma$-rays are potential candidates for neutrino emission. While evidence from multi wavelength observations mostly favors leptonic emission~\cite{LSIsource,Hesspaper1}, the possibility that protons or nuclei are being accelerated in the jets of these binary systems cannot be ruled out. The observation of neutrino emission from these sources would provide clear evidence for the presence of a hadronic component. 

Neutrinos could be produced in these binary systems during interactions of accelerated protons in relativistic jets with the atmosphere of the binary star~\cite{bib:periodicmodel}. These jets are narrow and they precess with the same time period as the binary system. The neutrino flux at Earth from these sources depends upon the orientation of these jets with respect to the atmosphere of the massive star and our line of sight, and hence is expected to be high only during a narrow time window of the orbit. 

The list of sources considered and the motivations are explained in detail in Ref.~\cite{periodical}. Three new sources were added to this list for a total of 10 sources (see Table~\ref{tab:table_periodic}). In the northern sky a recently reported binary HESS J0632+057~\cite{Hesspaper1} was added, which is a variable point like source of VHE ($>100$ GeV) $\gamma$-rays located in the Galactic plane and is positionally coincident with a Be star. It also emits variably in the radio and X-ray domains and has been found to have a hard X-ray spectrum~\cite{Hesspaper1}. The periodicity of the X-ray emission is $\Omega=320 \pm 5$ days~[53]. Bearing a close resemblance to the source LS I +61 303, this source has now been confirmed to be a $\gamma$-ray binary~\cite{Hesspaper2}. Motivated by the increased sensitivity of IceCube to neutrino sources in the southern sky as a consequence of background rejection techniques introduced recently with the first year of data from the 
completed IC-86I configuration of the detector~\cite{jakerameez}, two sources in the southern sky were also added. LS 5039 is a High Mass X-Ray Binary which was also the first micro-quasar to be established as a high-energy $\gamma$-ray source~\cite{HessLS5039Paper}. It has been confirmed to have a period of $3.906 \pm 0.002$ days ~\cite{HessLS5039Paper}. The second new source in the southern sky is HESS J1018-589, a Gamma Ray Binary in the Carina arm region of the galaxy. This source position is coincident with 1FGL J1018.6 reported by the Fermi-LAT Collaboration ~\cite{FermiHESSJ1018Paper} and is a source of ($>100$ GeV) $\gamma$-rays. Its periodicity has been reported to be $16.58 \pm 0.02$ days ~\cite{HESSJ1018Paper}.

\subsection{Method}
\label{periodic_method}

The analysis method used here was previously applied to data from the IC-22 and IC-40 configurations of the IceCube detector, and is explained in detail in~\cite{periodical}. It is adapted from the method used in the All Sky Time Scan (Sec \ref{sec:All-Sky Time Scan}). Each event within the sample is assigned an event phase $\phi_i$ based on the arrival time of the event and the period of the source, known mainly from optical measurements. Consequently, signal events coming from different periods of a periodic signal are assigned similar values of $\phi_i$.

The time dependence of the signal PDF $\mathcal{T}^{\mathrm{signal}}_{i}$ in equation~(\ref{eq:signalpdf}) is then replaced with a phase dependence $\mathcal{P}^{\mathrm{signal}}_{i}$ of the form:

\begin{equation}
    \mathcal{P}^{\mathrm{signal}}_{i}=\frac{1}{\sqrt{2\pi}\sigma_{\mathrm{\Phi}}}exp\left(-\frac{(\phi_i-\Phi)^2}{2\sigma^2_{\mathrm{\Phi}}}\right),
    \label{eq:time_pdf_allSkyPer}
\end{equation}

\noindent where $\phi_i$ is the arrival phase of the $i^{\mathrm{th}}$ event and $\Phi$ and $\sigma_{\mathrm{\Phi}}$ are the position of the peak signal in phase and the width respectively. These two parameters are fitted to maximize the likelihood. Two more free parameters were part of the fit, the number of signal events $n_{\mathrm{s}}$ and the spectral index $\gamma$. Thus the method searches for a statistically significant clustering of high energy events not only in space, but also in phase.

In~\cite{periodical} this search was constrained to look for a flare of duration larger than $0.02 \times \Omega$, where $\Omega$ is the period. This was done in order to avoid the fact that the minimizer tends to prefer shorter flares given a limited time window. The same marginalization term described previously for the ``All-Sky Time Scan'' was used in this search to prevent very short flares from dominating the significance, while still allowing a search for short flares which are of physical interest. The test statistic for this search including this modification is thus:

\begin{equation}
    TS = -2 \log\Big[\frac{1}{\sqrt{2\pi}\hat{\sigma}_{\mathrm{\Phi}}}\times\frac{\mathcal{L}(n_{\mathrm{s}}=0)}{\mathcal{L}(\hat{n}_{\mathrm{s}},\hat{\gamma}_{\mathrm{s}},\hat{\sigma}_{\mathrm{\Phi}},\hat{\Phi})}\Big],
    \label{eq:ts_allSkyPer}
\end{equation}

\noindent where $\hat{n}_{\mathrm{s}},\hat{\gamma}_{\mathrm{s}},\hat{\sigma}_{\mathrm{\Phi}},\hat{\Phi}$ are the best-fit values and $\frac{1}{\sqrt{2\pi}\hat{\sigma}_{\mathrm{\Phi}}}$ is the marginalization term.

\begin{figure}[h!]
  \centering
  \includegraphics[width=5.in]{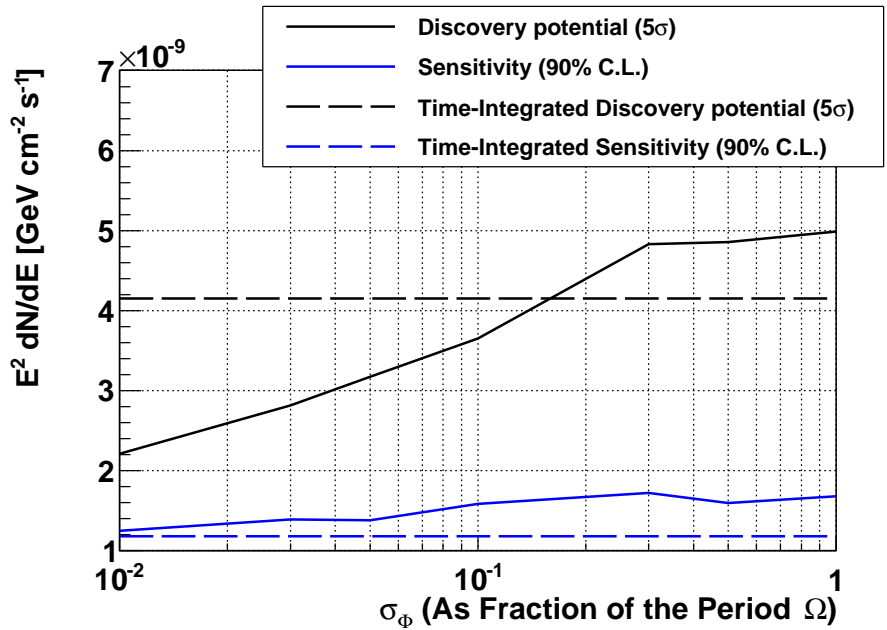}
  \caption{Discovery potential and sensitivity for four years of data (IC-40, IC-79, IC-59 and IC-86I) for periodic flares of varying width $\sigma$ (as a fraction of the total period), in terms of flux, for the source GRO J0422+32. The vertical axis denotes the mean flux over the period.}
  \label{fig:Periodic_Discpot}
\end{figure}

The post-trial p-value of the most significant observation was estimated by calculating the fraction of time-scrambled datasets in which the most significant fluctuation observed among the ten sources was more significant than that which was observed in non-scrambled data. As can be seen in Figure~\ref{fig:Periodic_Discpot}, the search is more sensitive to flares of very short duration measured as a fraction of the period of the system.

\begin{figure}[h!]
        \centering
        \includegraphics[width=5.in]{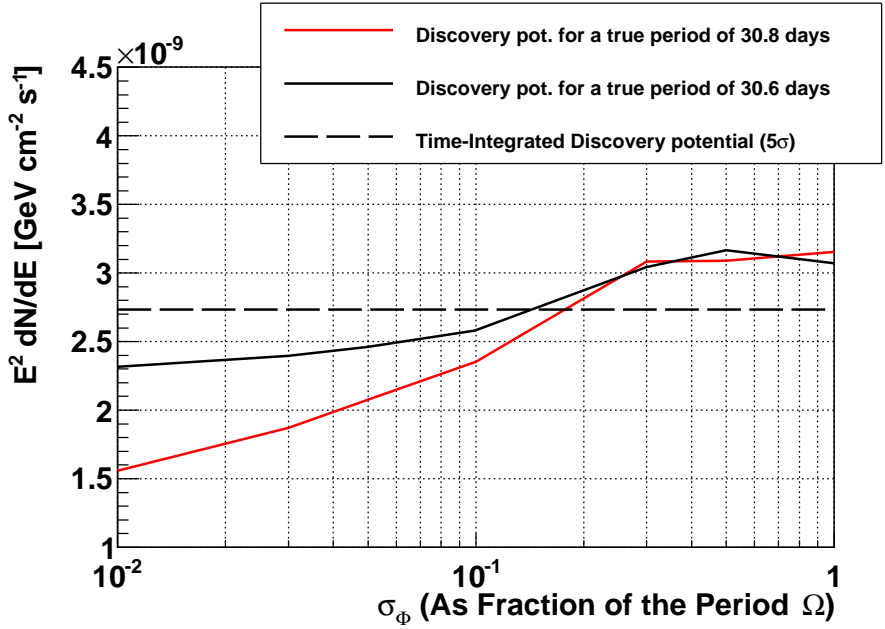}
        \caption{Impact of the uncertainty in the period of the source GRS 1915+105 for four years of data (IC-86I, IC-79, IC-59 and IC-40). The reported period is $30.8\pm0.2$ days. While the median value of 30.8 days is used to convert event times to phase, if the true period happens to be 30.6, the $5\sigma$ discovery potential from this search for narrow flares is still better than that of the time integrated search.}
        \label{fig:Periodic_Spreading}
    \end{figure}

As the considered periodic sources are not expected to change their long term behavior the performance of this search improves as more data are included. For the source GRS1915+105, which has large relative uncertainty on the measured period, converting the event times into phases using the reported central value may lead to a smearing out of the actual flare if it happened at a different time within the error.
Since the signal is modeled as a Gaussian flare in phase~\cite{periodical} and is more sensitive to narrower flares (Figure~\ref{fig:Periodic_Discpot}), this could negatively impact the sensitivities and the discovery potentials. The effect is proportional to the relative error on the period times the number of periods within the livetime and is hence most severe for GRS1915+105. The impact of the period uncertainty can be estimated by recalculating the discovery potential assuming the true source period differs from the central by the reported uncertainty. In Figure~\ref{fig:Periodic_Spreading} the result of this test is shown, demonstrating that this effect is not sufficient to negate the comparative advantage this search has over time integrated searches.

\subsection{Results}
\label{per_rez}
\begin{figure}[h!]
        \centering
        \includegraphics[width=5.in]{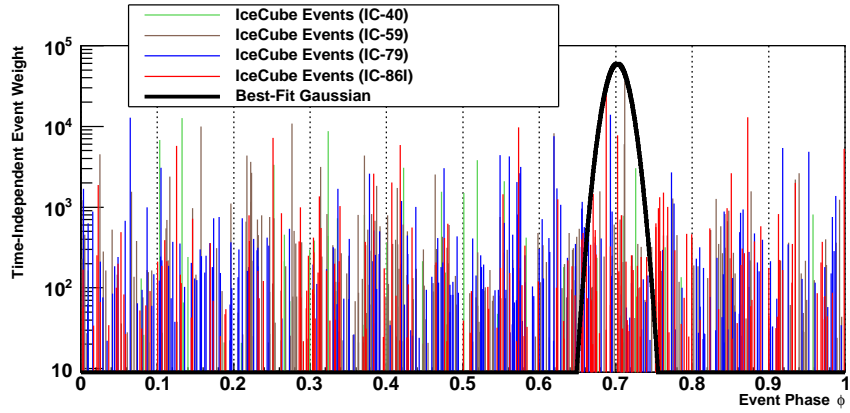}
        \caption{Events from the direction of HESS J0632+057 from which the most significant fluctuation from the background-only hypothesis was observed. The fitted flare at a phase of 0.7 with width 3.84 days is also shown.}
        \label{fig:HESSJ0632TimeStuff}
    \end{figure}

The results of the periodic analysis for each of the selected sources are given in Table~\ref{tab:table_periodic}. The most significant observation was from the source HESS J0632+057 with a pre-trial probability of 8.67\%. This Gaussian fitted flare was observed at a phase of 0.702 with a width of $\sigma_{\mathrm{\Phi}}=0.012$ in terms of the fraction of the period. Figure~\ref{fig:HESSJ0632TimeStuff} shows the fitted Gaussian and the IceCube event weights $w_i$ defined in equation~(\ref{eq:signalweight}). Cyg X-1, Cyg X-3 and and GRS 1915+105 were observed to have flares of probability 0.45, 0.34 and 0.32 respectively. All other sources produced under-fluctuations indicating that the number of events in the direction of the source was less than or equal to the number expected from background-only. The post-trial probability of the fluctuation from HESS J0632+057 was found to be 44.3\%, making the observation compatible with the background-only hypothesis.

\clearpage
\pagestyle{empty}
\newgeometry{top=3cm}
\begin{sidewaystable}[!h]
\begin{adjustwidth}{-1in}{-1in}
\centering
\begin{tabular}{|l|l|r|r|r|r|r|}
\toprule
\head{2.6cm}{Source [reference]} &\head{2.6cm}{Period(days)} & \head{1.3cm}{p-value} & \head{1.6cm}{flare duration} & \head{1.1cm}{phase} & \head{4.1cm}{Time dependent \newline90\% C.L. Upper Limit (GeV$^{-1}$cm$^{-2}$s$^{-1}$)} & \head{4.1cm}{Time integrated \newline90\% C.L. Upper Limit (GeV$^{-1}$cm$^{-2}$s$^{-1}$)} \\
\midrule
Cygnus X-1 \cite{Cygx1source} &  5.599829$\pm$0.000016 & 0.45 & 0.016 & 0.81 & $2.31\times 10^{-10}$ & $2.33\times 10^{-9}$\\
Cygnus X-3 \cite{bib:Fermi_cygnus} &  0.199679$\pm$0.000003 & 0.34 & 0.05 & 0.080 & $5.05\times 10^{-10}$ & $1.70\times 10^{-9}$\\
GRO J0422+32 \cite{GROsource} &  0.212140$\pm$0.000003 & - & - & - & - & $1.78\times 10^{-9}$ \\
GRS 1915+105 \cite{GRSsource} &  30.8$\pm$0.2 & 0.31 & 0.17 & 0.28 & $3.33\times 10^{-10}$ & $1.18\times 10^{-9}$\\
LSI + 61 303 \cite{LSIsource} &  26.496$\pm$0.0028 & - & - & -& - & $1.95\times 10^{-9}$\\
SS 433 \cite{SS4source} &  13.08227$\pm$0.00008 & - & - & - & - & $6.5\times 10^{-10}$\\
XTE J1118+480 \cite{XTEsource} &  0.1699339$\pm$0.0000002 & - & - & -& & $1.21\times 10^{-9}$\\
HESS J0632+057 \cite{Hessperiodpaper} &  320$\pm$5 & 0.087 & 0.0127 & 0.70 & $4.82\times 10^{-10}$ & $1.37\times 10^{-9}$\\
LS 5039 \cite{HessLS5039Paper} & 3.906 $\pm$ 0.002 & - & - & - & - & $5.24\times 10^{-9}$\\
HESS J1018-589 \cite{HESSJ1018Paper} & 16.58 $\pm$ 0.02 & - & - & - & - & $9.21\times 10^{-9}$\\
\bottomrule
\end{tabular}
\end{adjustwidth}
\caption{Candidate sources for the Periodic Neutrino emission search. The p-values are pre-trial. The flare-durations given are the widths of the the fitted gaussian flares, as fractions of the total periods. The time dependent upper limits are the normalization for an $E^{-2}$ power law flux with with time dependence corresponding to the fitted signal in likelihood, and is hence provided only for sources that were positive fluctuations. Time integrated Upper Limits~\cite{jakerameez} are a factor of $3 - 8$ times lower than reported by analyses performed on data from the IC-22 and IC-40 configurations of the detector \cite{periodical} and the time dependent upper limits have improved similarly, where comparable. }
\label{tab:table_periodic}
\end{sidewaystable}
\restoregeometry
\clearpage
\pagestyle{plain}

\clearpage
\section{Conclusions}
\label{sec:Conclusions}
Searches described within this paper have found no evidence for the existence of flaring or periodic neutrino sources. Both the untriggered search looking for neutrino flares anywhere in the sky and the triggered search looking for neutrino flares coinciding with $\gamma$-ray flares reported by the Fermi LAT returned results consistent with the background-only hypothesis. No evidence has been found for periodic neutrino emission by binary systems either. These analyses include data from the first year of operation of the completed IceCube detector, taken between May 2011 and May 2012. IceCube will continue to run in this configuration for the foreseeable future and the resultant signal build-up will increase the sensitivity to steady point sources of neutrinos \cite{jakerameez}. A similar improvement is expected in the sensitivity of the detector towards periodic neutrino signals. For flare searches, on the other hand, additional years of data with the full detector improve 
the chances to see rarer (rather than weaker) neutrino emission from outbursts which we have not been fortunate enough to witness yet.

In this paper we have demonstrated the viability of long term monitoring of sources of interest triggered by multiwavelength information from other experiments. As the detector matures, detector operations, data acquisition and processing will become more automated and we will soon be able to carry out this monitoring near-realtime - reducing the time delay between the trigger and the results.

\section*{Acknowledgements}
We acknowledge the support from the following agencies: U.S. National Science Foundation-Office of Polar Programs, U.S. National Science Foundation-Physics Division, University of Wisconsin Alumni Research Foundation, the Grid Laboratory Of Wisconsin (GLOW) grid infrastructure at the University of Wisconsin - Madison, the Open Science Grid (OSG) grid infrastructure; U.S. Department of Energy, and National Energy Research Scientific Computing Center, the Louisiana Optical Network Initiative (LONI) grid computing resources; Natural Sciences and Engineering Research Council of Canada, WestGrid and Compute/Calcul Canada; Swedish Research Council, Swedish Polar Research Secretariat, Swedish National Infrastructure for Computing (SNIC), and Knut and Alice Wallenberg Foundation, Sweden; German Ministry for Education and Research (BMBF), Deutsche Forschungsgemeinschaft (DFG), Helmholtz Alliance for Astroparticle Physics (HAP), Research Department of Plasmas with Complex Interactions (Bochum), Germany; Fund for Scientific Research (FNRS-FWO), FWO Odysseus programme, Flanders Institute to encourage scientific and technological research in industry (IWT), Belgian Federal Science Policy Office (Belspo); University of Oxford, United Kingdom; Marsden Fund, New Zealand; Australian Research Council; Japan Society for Promotion of Science (JSPS); the Swiss National Science Foundation (SNSF), Switzerland; National Research Foundation of Korea (NRF); Danish National Research Foundation, Denmark (DNRF).

\bibliographystyle{model1a-num-names}
\bibliography{<your-bib-database>}

\end{document}